\documentclass[a4paper,english,aps,prb,twocolumn]{revtex4}

\makeatletter

\providecommand{\tabularnewline}{\\}

\usepackage{color}
\usepackage{amsfonts}
\usepackage{amsmath}
\usepackage{amssymb}
\usepackage{graphicx}%
\usepackage{babel}
\makeatother
\begin{document}

\title{Dependence of the superconducting critical temperature on the number
of layers\\in homologous series of high-$T_c$ cuprates}

\author{T. A. Zaleski and T. K. Kope\'{c}}

\affiliation{Institute of Low Temperature and Structure Research, Polish Academy
of Sciences\\P.O. Box 1410, 50-950 Wroc\l aw, Poland}

\begin{abstract}
We study a model of $n$-layer high-temperature cuprates of homologous
series like \hbox{HgBa$_2$Ca$_{n-1}$Cu$_n$O$_{2+2n+\delta }$} to
explain the dependence of the critical temperature $T_{c}\left(n\right)$
on the number $n$ of \hbox{Cu-O} planes in the elementary cell.
Focusing on the description of the high-temperature superconducting
system in terms of the collective phase variables, we have considered
a semi-microscopic anisotropic three-dimensional vector XY model of
stacked copper-oxide layers with adjustable parameters representing
microscopic in-plane and out-of-plane phase stiffnesses. The model
captures the layered composition along $c$-axis of homologous series
and goes beyond the phenomenological Lawrence-Doniach model for layered
superconductors. Implementing the spherical closure relation for vector
variables we have solved the phase XY model exactly with the help
of transfer matrix method and calculated $T_{c}\left(n\right)$ for
arbitrary block size $n$, elucidating the role of the $c$-axis anisotropy
and its influence on the critical temperature. Furthermore, we accommodate
inhomogeneous charge distribution among planes characterized by the
charge imbalance coefficient $R$ being the function of number of
layers $n$. By making a physically justified assumption regarding
the doping dependence of the microscopic phase stiffnesses, we have
calculated the values of parameter $R$ as a function of block size
$n$ in good agreement with the nuclear magnetic resonance data of
carrier distribution in multilayered high-$T_{c}$ cuprates. 
\end{abstract}


\maketitle

\section{Introduction}

Characteristic common structure of high-temperature superconducting
cuprates has been puzzling researchers since the discovery of the
phenomenon by Bednorz and M\"{u}ller.\cite{MullerBednorz} It is
now accepted that all relevant carriers of both spin and electricity
derive from the hybridized antibonding orbitals in the \hbox{Cu-O}
planes.\cite{AndersonDogmas} Obviously, the interlayer structure
cannot be ignored.\cite{LeggettScience} In the normal state, due
to huge anisotropy, coherent transport in the direction perpendicular
to \hbox{Cu-O} planes is blocked. However, it is observed that $T_{c}$
is strongly dependent on the interlayer structure, which means that
three-dimensional (3D) coupling of planes must play an important role
in the onset of superconductivity.\cite{Intercalation} It suggested
to apply the Lawrence and Doniach model (LD) to cuprates, which originally
was proposed for layered superconductors and is based on the Ginzburg-Landau
description with neighboring layers coupled by the Josephson tunneling.\cite{LawrenceDoniach}
This approach was pursued to study Josephson plasma resonance,\cite{JosephsonPlasma}
anisotropy of critical fields $H_{c1}$ and $H_{c2}$,\cite{FieldAnisotropy,FieldAnisotropy2}
vortex matter behavior.\cite{Vortex1} However, the biggest disadvantage
of the LD model is that it is based on mean-field approximation, which
neglects the role of phase fluctuations, the latter playing a profound
role in cuprates.\cite{EmeryKivelson}

There is a significant evidence that in cuprates the \hbox{Cu-O}
planes are Josephson coupled and the anisotropy of $c$-axis to $ab$-planes
penetration depth can vary from 5 to even 250 in various high-$T_{c}$
compounds. This gave rise to \emph{interplane} theories e.g. interlayer
tunneling model (ILT) by Anderson,\cite{ILTAnderson} which assumed
that carriers were incoherent between layers in the normal phase,
however coherent transport of Cooper pairs in a form of Josephson
tunneling in superconducting state was allowed. Unfortunately, the
original version of the ILT theory by Anderson is no longer considered
as a viable model of high-\textbf{$T_{c}$} materials, since it provides
no more than 1\% of the condensation energy in certain cuprates.\cite{Moler,Tsvetkov}
However, the Josephson coupling between \hbox{Cu-O} planes is still
believed to play an important role in cuprate oxides.\cite{Schneider}
In the process of synthesis of high-$T_{c}$ materials it is possible
to control (to some extent) the distances between copper-oxide planes
or the content of the regions that separate them. It allows to study,
how such changes influence observed properties of these compounds
like critical temperature or electrical resistivity and elucidate
the role of the inter-layer coupling.\cite{Choi,Xiang} In mono- or
bi-layered high-$T_{c}$ materials (like LSCO or YBCO compounds),
all copper-oxide planes are equivalent, which means that charge carriers
are distributed homogeneously among copper-oxide planes. However,
\emph{multi-layered} high-$T_{c}$ materials are characterized by
the existence of crystalographically inequivalent outer (OP) and inner
(IP) \hbox{Cu-O} planes, which give rise to inhomogeneous carrier
distribution among them. This, in turn, may have a profound influence
on the critical temperature, which is know to be strongly doping dependent.
For example, high-temperature superconducting homologous series, like
\hbox{HgBa$_2$Ca$_{n-1}$Cu$_n$O$_{2+2n+\delta }$}, are materials
sharing the same charge reservoir, but differing by the number $n$
of \hbox{Cu-O} planes in the unit cell.\cite{MultiLayer1,HgBCCOLatticeParams}
They also have the same characteristic dependence of the critical
temperature as a function of $n$ with a maximum for $n=3$ or $n=4$.
Although, these groups of cuprates have been extensively studied,
the origin of this extremum is still not fully understood.\cite{ILT,Leggett,ILT2} 

Recently, Chakravarty et al. have argued that the existence of the
competing order parameter that is nucleated by charge imbalance is
ultimately responsible for the critical temperature downturn with
$n$.\cite{Chakravarty} They present a mean-field theory for a single
block of $n$ superconducting layers, however neglecting the \emph{inter-block}
coupling. Moreover, they use phenomenological Ginzburg-Landau theory
at \emph{zero-temperature} to anticipate the values of the critical
temperature $T_{c}\left(n\right)$, based on the magnitude of the
\hbox{$T=0$} superconducting order parameter. Charge imbalance is
introduced as a competing order parameter, e.g. $d$-density wave.\cite{Chakravarty2}
It depresses the critical temperature for $n>2$, thus experimentally
observed shape of $T_{c}\left(n\right)$ is obtained. However, as
the authors note, the phase fluctuations, which are not taken into
account, might depress $T_{c}$ for $n=1$ even more than for $n=3$.
Although, it should correct the results for $1<n<3$, it is not obvious,
whether the maximum of $T_{c}\left(n\right)$ can be observed at all. 

It is well known that binding of electrons into pairs is essential
in forming the superconducting state, however, its remarkable properties
-- zero resistance and Meissner effect -- require the \emph{phase
coherence} among the pairs as well. In conventional BCS superconductors
the phase order is unimportant for determining of the value of the
transition temperature $T_{c}$. However, in superconductors with
low carrier density such as high-$T_{c}$ oxide superconductors, phase
fluctuations may have a profound influence on low temperature properties.\cite{EmeryKivelson}
In particular, for cuprate superconductors, the conventional ordering
of binding and phase stiffness energies appears to be reversed. Thus,
the issue how pairing and phase correlations develop is a central
problem of high-$T_{c}$ superconductivity. The measurements of the
frequency dependent conductivity, in the frequency range 100-600GHz,
shows that phase correlations indeed persist above $T_{c}$, where
the phase dynamics is governed by the bare microscopic phase stiffnesses.\cite{VanishingPhaseCoherence} 

In the present paper we propose a \emph{semi-microscopic} model of
$n$-layer cuprates, which is founded on \emph{microscopic phase stiffnesses}
that set the characteristic energy scales: in-plane $J_{\|}$, inter-plane
(in-block) $J_{\bot}$ and inter-block $J_{\bot}^{'}$ couplings.
We employ an approach that goes beyond the mean field level and is
able to capture both the effects of phase fluctuations and huge $c$-axis
anisotropy on the superconducting phase transition. We show that the
depression of the critical temperature for $n>3$ can be obtained
naturally without invoking any competing (hidden) order. Since superconducting
carriers are located in \hbox{Cu-O} planes, the in-plane phase stiffness
must naturally be doping $\delta$ dependent. As this dependence can
be simply deduced from the established $T_{c}-\delta$ phase diagram
for cuprates, introduction of charge imbalance via the doping dependence
of microscopic in-plane phase stiffness allows us for explanation
of $T_{c}\left(n\right)$ as a function of block layers $n$ in perfect
agreement with experimental findings.\cite{MultiLayer1} 

The outline of the reminder of the paper is as follows. In Section
II we analyze a single-layer system. We construct an anisotropic three-dimensional
XY model, which will be solved in the spherical approximation to explicate
the dependence of critical temperature on the anisotropy $\eta=J_{\bot}/J_{\|}$
along $c$-axis. Subsequently, in Section III we elaborate on a model
of stacked $n$-block layers relevant for high-temperature superconducting
homologous series. We map the system onto a spherical XY model, which
is solved exactly for arbitrary $n$ with the help of the transfer
matrix method. In Section IV we determine the critical temperature
$T_{c}\left(n\right)$ as a function of number of layers $n$. In
Section V we incorporate the inhomogeneous charge distribution in
homologous series by introducing doping dependent in-plane phase stiffnesses
$J_{\|}\left(\delta\right)$ for the separate $n$ layers in the block
structure. We found that the maximum of the phase coherent superconducting
transition will be a bell-shaped curve as a function of $n$ with
a maximum for $n=3$ for appropriate values of outer ($\delta_{op}$)
and inner ($\delta_{ip}$) planes dopings within a block. Finally,
in Section VI we summarize the conclusions to be drawn from our work.
Some supplementary material is relegated to the Appendix.

\section{Single layer cuprates\label{sec:Spherical-model}}

\begin{figure}
\includegraphics[%
  scale=0.45]{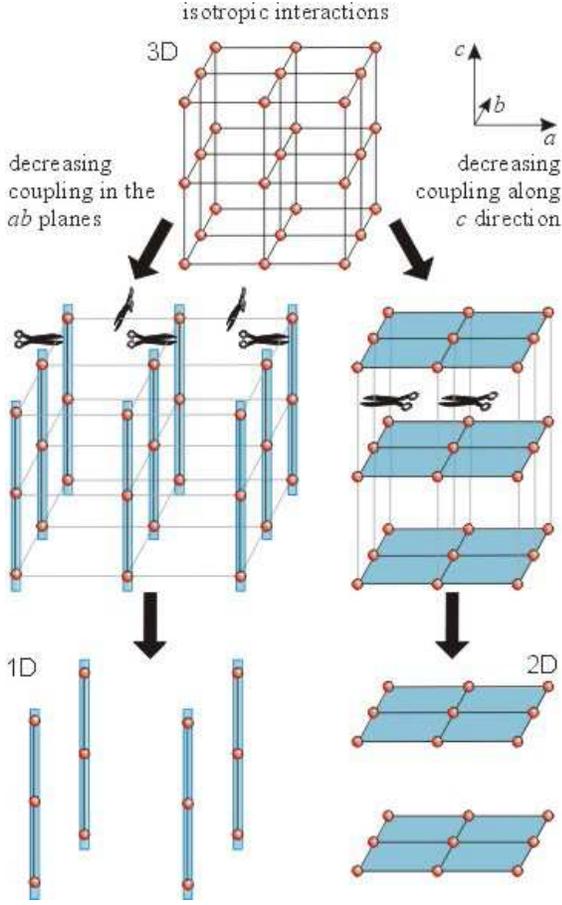}

\caption{(Color online) Pictorial representation of the various crossovers
due to increased spatial anisotropy between in-plane and inter-plane
couplings: 3D$\rightarrow$2D (isotropic interactions to uncoupled
planes) and 3D$\rightarrow$1D (isotropic interactions to uncoupled
chains).}

\label{Fig312DHowTo}
\end{figure}

In underdoped high-temperature superconductors, two temperature scales
of short-length pairing correlations and long-range superconducting
order seem to be well separated \cite{EmeryKivelson}. We consider
the situation, in which local superconducting pair correlations are
established and the relevant degrees of freedom are represented by
phase factors $0\le\varphi_{\ell}\left(\mathbf{r}_{i}\right)<2\pi$,
where $\mathbf{r}_{i}$ numbers lattice sites within $\ell$-th $ab$
plane. The system becomes superconducting once U(1) symmetry group
governing the $\varphi_{\ell}\left(\mathbf{r}_{i}\right)$ factors
is spontaneously broken and the non-zero value of $\left\langle e^{i\varphi_{\ell}\left(\mathbf{r}_{i}\right)}\right\rangle $
appears signaling the long-range phase order. The dependence of critical
temperature on the spatial $c$-axis anisotropy is of paramount importance
for cuprate systems, in which two-dimensional physics predominate.
First, we consider a single-layer system described by the following
Hamiltonian:\begin{eqnarray}
H\left[\varphi\right] & = & -J_{\|}\sum_{\ell}\sum_{i<j}\cos\left[\varphi_{\ell}\left(\mathbf{r}_{i}\right)-\varphi_{\ell}\left(\mathbf{r}_{j}\right)\right]\nonumber \\
 &  & -J_{\bot}\sum_{\ell}\sum_{i}\cos\left[\varphi_{\ell}\left(\mathbf{r}_{i}\right)-\varphi_{\ell+1}\left(\mathbf{r}_{i}\right)\right],\end{eqnarray}
where $J_{\|}>0$ and $J_{\bot}>0$ are in-plane and inter-plane microscopic
phase stiffnesses, respectively. The factors $\varphi_{\ell}\left(\mathbf{r}_{i}\right)$
are placed in sites of a three-dimensional cubic lattice, where $i=1,...,N_{\|}$
with $N_{\|}$ being the number of sites in a plane, $\ell=1,...,N_{\bot}$,
where $N_{\bot}$ denotes the number of layers and $N=N_{\|}N_{\bot}$
is the total number of sites. The partition function of the system
reads:\begin{equation}
Z=\int_{0}^{2\pi}\prod_{\ell,i}d\varphi_{\ell}\left(\mathbf{r}_{i}\right)e^{-\beta H\left[\varphi\right]},\label{eq:PartFunctionFirst}\end{equation}
where $\beta=1/k_{B}T$ with $T$ being the temperature. Introducing
two-dimensional vectors $\mathbf{S}_{\ell}\left(\mathbf{r}_{i}\right)=\left[S_{x\ell}\left(\mathbf{r}_{i}\right),\, S_{y\ell}\left(\mathbf{r}_{i}\right)\right]$
of the unit length $\mathbf{S}_{\ell}^{2}\left(\mathbf{r}_{i}\right)=S_{x\ell}^{2}\left(\mathbf{r}_{i}\right)+S_{y\ell}^{2}\left(\mathbf{r}_{i}\right)=1$
defined by \begin{equation}
\mathbf{S}_{\ell}\left(\mathbf{r}_{i}\right)=\left[\cos\varphi_{\ell}\left(\mathbf{r}_{i}\right),\sin\varphi_{\ell}\left(\mathbf{r}_{i}\right)\right],\label{SVectors}\end{equation}
 the Hamiltonian can be written in the XY-model form:\begin{eqnarray}
H\left[\mathbf{S}\right] & = & -J_{\|}\sum_{\ell}\sum_{i<j}\mathbf{S}_{\ell}\left(\mathbf{r}_{i}\right)\cdot\mathbf{S}_{\ell}\left(\mathbf{r}_{j}\right)\nonumber \\
 &  & -J_{\bot}\sum_{\ell}\sum_{i}\mathbf{S}_{\ell}\left(\mathbf{r}_{i}\right)\cdot\mathbf{S}_{\ell+1}\left(\mathbf{r}_{i}\right).\label{SphericalHamiltonian}\end{eqnarray}
In terms of the vector variables the partition function in Eq. (\ref{eq:PartFunctionFirst})
becomes:\begin{eqnarray}
Z & = & \int_{-\infty}^{+\infty}\prod_{i,\ell}\left\{ d^{2}\mathbf{S}_{\ell}\left(\mathbf{r}_{i}\right)\delta\left[\mathbf{S}_{\ell}^{2}\left(\mathbf{r}_{i}\right)-1\right]\right\} e^{-\beta H\left[\mathbf{S}\right]},\label{eq:SinglePartFunction}\end{eqnarray}
where $d^{2}\mathbf{S}_{\ell}\left(\mathbf{r}_{i}\right)\equiv dS_{x\ell}\left(\mathbf{r}_{i}\right)dS_{y\ell}\left(\mathbf{r}_{i}\right)$
and the Dirac-$\delta$ function $\delta\left[\mathbf{S}_{\ell}^{2}\left(\mathbf{r}_{i}\right)-1\right]$
assures that the integration over $\mathbf{S}_{\ell}\left(\mathbf{r}_{i}\right)$
variables runs only over the values, which satisfy the unit length
condition $\mathbf{S}_{\ell}^{2}\left(\mathbf{r}_{i}\right)=1$. Unfortunately,
the partition function in Eq. (\ref{eq:SinglePartFunction}) cannot
be evaluated exactly. However, replacing of the rigid length constraint
in Eq. (\ref{eq:SinglePartFunction}) by a weaker spherical closure
relation\cite{SphericalModel,SphericalComment}\begin{equation}
\delta\left[\mathbf{S}_{\ell}^{2}\left(\mathbf{r}_{i}\right)-1\right]\,\,\,\rightarrow\,\,\,\delta\left[\frac{1}{N}\sum_{i,\ell}\mathbf{S}_{\ell}^{2}\left(\mathbf{r}_{i}\right)-1\right].\label{SphericalConstraint}\end{equation}
renders the model in Eq. (\ref{SphericalHamiltonian}) exactly solvable.
The relation in Eq. (\ref{SphericalConstraint}) means that the unit
length of the $\mathbf{S}_{\ell}\left(\mathbf{r}_{i}\right)$ vectors
is maintained on average. The spherical constraint can be conveniently
implemented with the help of Dirac-$\delta$ function and the partition
function in Eq. (\ref{eq:SinglePartFunction}) becomes:\begin{equation}
Z=\int_{-\infty}^{+\infty}\prod_{i,\ell}d^{2}\mathbf{S}_{\ell}\left(\mathbf{r}_{i}\right)\delta\left[\sum_{i,\ell}\mathbf{S}_{\ell}^{2}\left(\mathbf{r}_{i}\right)-N\right]e^{-\beta H\left[\mathbf{S}\right]}.\label{eq:SinglePartSpherical}\end{equation}
Furthermore, the $\delta$-function can be conveniently represented
in a spectral form by the appropriate integral $\delta\left(x\right)=\int_{-\infty}^{\infty}d\zeta/2\pi i\exp\left(-\zeta x\right)$,
which introduces the Lagrange multiplier $\zeta$. Consequently, Eq.
(\ref{eq:SinglePartSpherical}) can be written as: \begin{equation}
Z=\int_{-\infty}^{+\infty}\frac{d\zeta}{2\pi i}\exp\left[-N\beta f\left(\zeta\right)\right],\label{eq:SinglePartFinal}\end{equation}
 where the free energy per site is given by: \begin{eqnarray}
f\left(\zeta\right) & = & -\frac{\zeta}{\beta}-\frac{1}{N\beta}\ln\int_{-\infty}^{+\infty}\prod_{i,\ell}d^{2}\mathbf{S}_{\ell}\left(\mathbf{r}_{i}\right)\nonumber \\
 & \times & \exp\left\{ -\zeta\sum_{i,\ell}\mathbf{S}_{\ell}^{2}\left(\mathbf{r}_{i}\right)-\beta H\left[\mathbf{S}\right]\right\} .\label{eq:FreeEnergy}\end{eqnarray}
 In the thermodynamic limit $N\rightarrow\infty$ the dominant contribution
to the integral in Eq. (\ref{eq:SinglePartFinal}) comes from the
saddle point $\zeta=\zeta_{0}$ of $f\left(\zeta\right)$ with value
of $\zeta_{0}$ determined from the condition: \begin{equation}
\left.\frac{\partial f\left(\zeta\right)}{\partial\zeta}\right|_{\zeta=\zeta_{0}}=0.\label{Spherical_SaddlePointEquation}\end{equation}
 To evaluate the free energy of the system $f=f\left(\zeta_{0}\right)$
we diagonalize the Hamiltonian from Eq. (\ref{SphericalHamiltonian})
by introducing Fourier transform of $\mathbf{S}_{\ell}\left(\mathbf{r}_{i}\right)$
variables:\begin{equation}
\mathbf{S}_{\ell}\left(\mathbf{r}_{i}\right)=\frac{1}{N_{\|}N_{\bot}}\sum_{\mathbf{k},q_{z}}\mathbf{S}_{\mathbf{k},q_{z}}e^{-i\left(\mathbf{kr}_{i}+q_{z}c\ell\right)},\label{Spherical_Fourier}\end{equation}
where $\mathbf{k}=\left(k_{x},\, k_{y}\right)$ with $-\frac{\pi}{a}<k_{x},\, k_{y}<\frac{\pi}{a}$
are wave vectors within \textbf{$ab$}-planes, while $-\frac{\pi}{c}<q_{z}<\frac{\pi}{c}$
is wave vector along \textbf{$c$}-axis. Here, $a$ and $c$ are lattice
constant within \textbf{$ab$}-plane and $c$-axis, respectively.
The Hamiltonian in Eq. (\ref{SphericalHamiltonian}) then transforms
to:\begin{eqnarray}
H & = & -\frac{1}{2N}\sum_{\mathbf{k},q_{z}}J\left(\mathbf{k},q_{z}\right)\mathbf{S}_{\mathbf{k},q_{z}}\cdot\mathbf{S}_{-\mathbf{k},-q_{z}},\end{eqnarray}
where for the Fourier-transformed interactions $J_{\|}$ and $J_{\bot}$
on the 3D anisotropic cubic lattice with nearest-neighbor coupling
we readily find:\begin{equation}
J\left(\mathbf{k},q_{z}\right)=2J_{\|}\left[\cos\left(ak_{x}\right)+\cos\left(ak_{y}\right)\right]+2J_{\bot}\cos\left(cq_{z}\right).\end{equation}
As a result, the integral in Eq.~(\ref{eq:FreeEnergy}) becomes Gaussian
and the free energy can be calculated:\begin{eqnarray}
f & = & -\frac{\zeta_{0}}{\beta}-\frac{1}{\beta N}\ln\int_{-\infty}^{+\infty}\prod_{\mathbf{k},q_{z}}d^{2}\mathbf{S}_{\mathbf{k},q_{z}}\nonumber \\
 & \times & \exp\left\{ -\frac{1}{N}\sum_{\mathbf{k,}q_{z}}\left[\zeta_{0}-\frac{1}{2}J\left(\mathbf{k},q_{z}\right)\right]\mathbf{S}_{\mathbf{k},q_{z}}^{*}\mathbf{S}_{\mathbf{k},q_{z}}\right\} ,\nonumber \\
 &  & d^{2}\mathbf{S}_{\mathbf{k},q_{z}}\equiv\frac{d\mathrm{Re}\mathbf{S}_{\mathbf{k},q_{z}}^{*}d\mathrm{Im}\mathbf{S}_{\mathbf{k},q_{z}}}{\pi}.\label{eq:SingleFreeEnergy3}\end{eqnarray}
The emergence of the critical point in the spherical model is signaled
by the condition\cite{SphericalModel} \begin{equation}
G^{-1}\left(\mathbf{k}=0,q_{z}=0\right)\equiv\zeta_{0}-\frac{1}{2}\beta J_{0}=0,\label{SphericalCorrelator}\end{equation}
which fixes the saddle-point value of the Lagrange multiplier $\zeta$
within the ordered phase:\begin{equation}
\zeta_{0}=\frac{1}{2}\beta J_{0},\end{equation}
where $J_{0}=J\left(\mathbf{k}=0,q_{z}=0\right)$. Here, $G\left(\mathbf{k},q_{z}\right)$
is the order parameter susceptibility defined by:\begin{equation}
G\left(\mathbf{k},q_{z}\right)=\left\langle \mathbf{S}_{\mathbf{k},q_{z}}\cdot\mathbf{S}_{-\mathbf{k},-q_{z}}\right\rangle ,\end{equation}
 with the statistical average:\begin{eqnarray}
\left\langle ...\right\rangle  & = & \frac{1}{Z}\int_{-\infty}^{+\infty}\prod_{i,\ell}d^{2}\mathbf{S}_{\ell}\left(\mathbf{r}_{i}\right)\nonumber \\
 &  & \times\delta\left[\sum_{i,\ell}\mathbf{S}_{\ell}^{2}\left(\mathbf{r}_{i}\right)-N\right]...e^{-\beta H\left[\mathbf{S}\right]}.\end{eqnarray}
 From Eqs. (\ref{Spherical_SaddlePointEquation}) and (\ref{eq:SingleFreeEnergy3})
we obtain the critical temperature in the form:\begin{eqnarray}
\beta_{c} & =\underset{N\rightarrow\infty}{\lim} & \frac{2}{N}\sum_{\mathbf{\mathbf{k}},q_{z}}\frac{1}{J_{0}-J\left(\mathbf{\mathbf{k}},q_{z}\right)}.\label{SphericalCritTemp}\end{eqnarray}

For one (1D) and two-dimensional (2D) systems with short-range interactions
the ordered phase exists only in the zero temperature limit, hence
for dimensions $d\le2$ the spherical model does not exhibit phase
transition with $T_{c}>0$, in agreement with Mermin-Wagner theorem.\cite{SphericalModel}
For $d=3$ the critical temperature is finite \begin{equation}
\frac{2k_{B}T_{c}}{J_{0}}\approx0.659,\end{equation}
 where the factor $2$ in front of $k_{B}T_{c}$ comes from the number
of components of the vector XY model.\cite{SphericalModel}

\begin{figure}
\includegraphics[%
  scale=0.4]{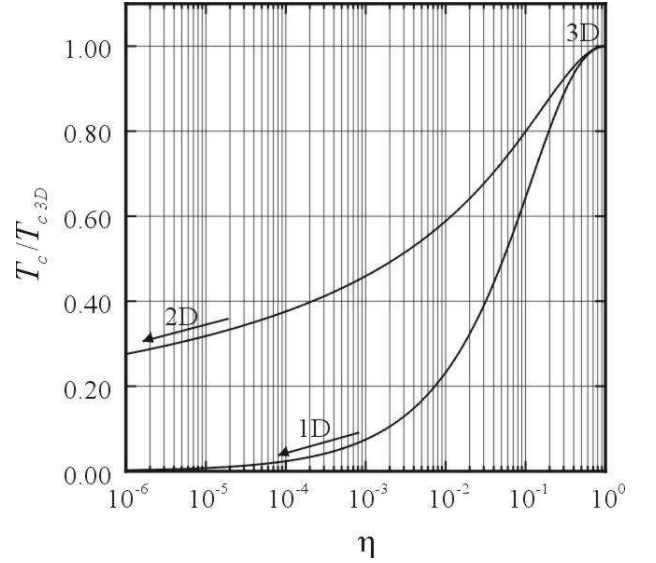}

\caption{Superconducting critical temperature $T_{c}$ as a function of anisotropy
$\eta$ for crossovers from isotropic 3D to 1D or 2D anisotropic systems
in logarithmic scale for $\eta$.}

\label{Fig312D}
\end{figure}

Introducing spatial anisotropy parameter $0\le\eta\le1$ (being the
ratio between $J_{\|}$ and $J_{\bot}$), $J\left(\mathbf{k},q_{z}\right)$
can be written in a form to accommodate 3D$\rightarrow$2D crossover
from the isotropic superconductor on the three-dimensional cubic lattice
($\eta=1$) to the uncoupled ($\eta=0$) copper-oxide planes (see,
Fig. \ref{Fig312D}): \begin{eqnarray}
J\left(\mathbf{k},q_{z}\right) & = & 2J\left[\cos\left(ak_{x}\right)+\cos\left(ak_{y}\right)+\eta\cos\left(cq_{z}\right)\right]\nonumber \\
 &  & J\equiv J_{\|},\quad\eta=J_{\bot}/J_{\|}.\end{eqnarray}
In a similar way we can study a 3D$\rightarrow$1D crossover from
the isotropic superconductor to the uncoupled one-dimensional chains:\begin{eqnarray}
J\left(\mathbf{k},q_{z}\right) & = & 2J\left\{ \eta\left[\cos\left(ak_{x}\right)+\cos\left(ak_{y}\right)\right]+\cos\left(cq_{z}\right)\right\} \nonumber \\
 &  & J\equiv J_{\bot},\quad\eta=J_{\|}/J_{\bot}.\end{eqnarray}
The critical temperature can be found for both cases from Eq. \ref{SphericalCritTemp}
by performing momentum integration. In the thermodynamic limit we
have:\begin{equation}
\underset{N\rightarrow\infty}{\mathrm{lim}}\frac{1}{N}\sum_{\mathbf{k},q_{z}}\dots=\int_{-\frac{\pi}{a}}^{\frac{\pi}{a}}\frac{d^{2}\mathbf{k}}{\left(2\pi/a\right)^{2}}\int_{-\frac{\pi}{c}}^{\frac{\pi}{c}}\frac{dq_{z}}{\left(2\pi/c\right)}\dots\,\,.\end{equation}
For 3D$\rightarrow$2D crossover it reads:\begin{eqnarray}
 &  & \frac{J_{0}}{k_{B}T_{c}\left(\eta\right)}=\frac{2\left(2+\eta\right)\left(2\sqrt{1+\eta}-\sqrt{4+2\eta}\right)}{\eta\pi^{2}}\nonumber \\
 &  & \times\mathbf{K}\left[\frac{1}{2\eta}\left(2\sqrt{1+\eta}-\sqrt{4+2\eta}\right)\left(\sqrt{4+2\eta}+2\right)\right]\nonumber \\
 &  & \times\mathbf{K}\left[\frac{1}{2\eta}\left(2\sqrt{1+\eta}-\sqrt{4+2\eta}\right)\left(\sqrt{4+2\eta}-2\right)\right],\nonumber \\
\label{eq:3D2D}\end{eqnarray}
while for 3D$\rightarrow$1D changeover the Eq. (\ref{SphericalCritTemp}):

\begin{eqnarray}
 &  & \frac{J_{0}}{k_{B}T_{c}\left(\eta\right)}=\frac{2\left(1+2\eta\right)\left(2\sqrt{1+\eta}-\sqrt{2-4\eta}\right)}{\pi^{2}\sqrt{\eta}}\nonumber \\
 &  & \,\,\times\mathbf{K}\left[2\left(\sqrt{\eta}-\sqrt{\frac{1}{2}+\eta}\right)\left(\sqrt{\frac{1}{2}+\eta}-\sqrt{1+\eta}\right)\right]\nonumber \\
 &  & \,\,\times\mathbf{K}\left[2\left(\sqrt{\eta}+\sqrt{\frac{1}{2}+\eta}\right)\left(\sqrt{\frac{1}{2}+\eta}-\sqrt{1+\eta}\right)\right],\nonumber \\
\label{eq:3D1D}\end{eqnarray}
where $\mathbf{K}\left(x\right)$ is the complete elliptic integral
of the first kind.\cite{EllipticIntegral} It is interesting, how
these $T_{c}\left(\eta\right)$ values behave in the limiting cases.
For the 3D$\rightarrow$2D crossover from Eq. (\ref{eq:3D2D}) we
obtain:\begin{equation}
\frac{k_{B}T_{c}}{J_{0}}\approx\left\{ \begin{array}{l}
1.319-0.0944\eta^{2}-0.0559\eta^{4}\,\,\mathrm{for\,\eta\rightarrow1}\\
{\displaystyle \frac{2\pi}{\ln\left(32/\eta\right)}-\frac{\pi\eta\left[\ln\left(32/\eta\right)+2\right]}{2\ln^{2}\left(32/\eta\right)}}\,\,\mathrm{for\,\eta\rightarrow0},\end{array}\right.\end{equation}
while for 3D$\rightarrow$1D changeover, Eq. (\ref{eq:3D1D}) it follows
that:\begin{equation}
\frac{k_{B}T_{c}}{J_{0}}\approx\left\{ \begin{array}{l}
1.319-0.0944\eta^{2}-0.106\eta^{4}\,\,\mathrm{for\,\eta\rightarrow1}\\
{\displaystyle \frac{\left(2+\sqrt{2}\right)\pi^{2}\sqrt{\eta}}{4\mathrm{\mathbf{K}}^{2}\left(3-2\sqrt{2}\right)}}\,\,\mathrm{for\,\eta\rightarrow0}.\end{array}\right.\end{equation}

The results are presented in Fig. \ref{Fig312D} and in Table \ref{Table312D}.
When the in-plane interactions are reduced, the critical temperature
drops quite rapidly as a function of $\eta$ compared to the isotropic
system and for $\eta=10^{-5}$ is less than 1\% of the isotropic value.
However, in the case of the reduced coupling along $c$-axis (weakly
coupled planes), even for anisotropy parameter as small as $\eta=10^{-6}$
the critical temperature is still about 30\% of the isotropic three-dimensional
value. This result is very important as it proves that even very small
inter-planar coupling (sometimes too small to be directly observed
experimentally) renders the phase transition in the 3D universality
class,\cite{Schneider} with the critical temperature decreasing slowly
like $T_{c}\sim1/\log\eta$. This result raises serious doubts whether
the outcome of intercalation experiments in cuprates,\cite{Choi}
which show small changes of the critical temperature with increase
of the inter-layer distance can be interpreted as an argument in favor
for a purely 2D nature of high-$T_{c}$ superconductivity.

\begin{table}
\begin{tabular}{|c|c|c|}
\hline 
$\hspace{1cm}\eta\hspace{1cm}$&
3D$\rightarrow$2D:~~$\frac{T_{c}\left(\eta\right)}{T_{c}\left(1\right)}$&
3D$\rightarrow$1D:~~$\frac{T_{c}\left(\eta\right)}{T_{c}\left(1\right)}$\tabularnewline
\hline
\hline 
1&
1&
1\tabularnewline
\hline 
0.1&
0.80&
0.64\tabularnewline
\hline 
0.01&
0.59&
0.23\tabularnewline
\hline 
0.001&
0.46&
0.074\tabularnewline
\hline 
0.0001&
0.38&
0.024\tabularnewline
\hline 
0.00001&
0.32&
0.0075\tabularnewline
\hline 
0.000001&
0.28&
0.0024\tabularnewline
\hline
\end{tabular}

\caption{Ratio of the superconducting critical temperature $T_{c}\left(\eta\right)$
to the isotropic three-dimensional value $T_{c}\left(1\right)$ for
several values of the anisotropy parameter $\eta$ facilitating the
crossovers from isotropic 3D to anisotropic 1D or 2D systems.}

\label{Table312D}
\end{table}

\section{Multi layered system}

\begin{figure}
\includegraphics[%
  scale=0.4]{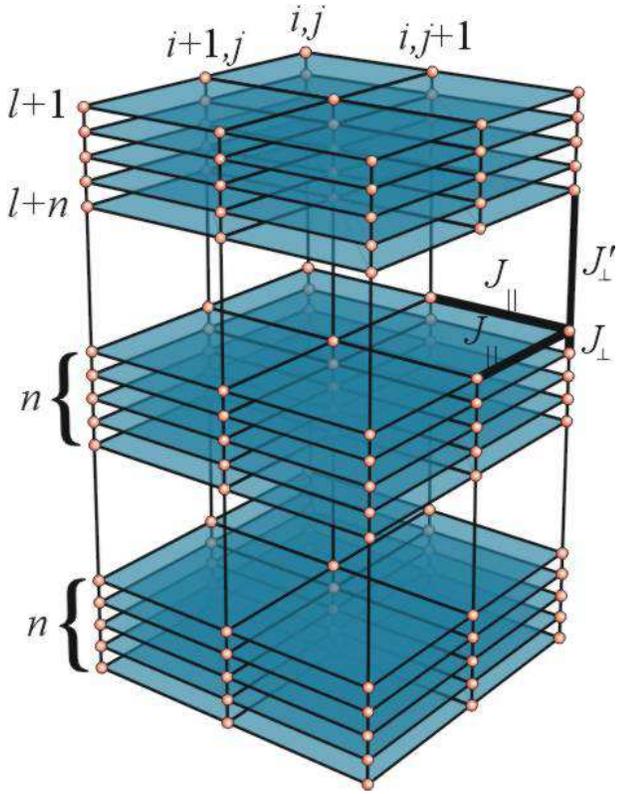}

\caption{(Color online) Structure of the multi-layer high-temperature superconductors
of homologous series of cuprates. Coupled \hbox{Cu-O} planes are
grouped in blocks of $n$ planes. The microscopic phase stiffnesses
$J_{\|}$, $J_{\bot}$ and $J_{\bot}^{'}$ are indicated by the thick
lines. }

\label{FigModel}
\end{figure}

To describe superconducting cuprates of homologous families, we modify
the model presented in Sec. \ref{sec:Spherical-model} to include
the presence of the $n$-layer block structure. To this end, we group
every $n$ consecutive layers in blocks. The intra-layer (in the $ab$
planes) interactions are given by $J_{\|}>0$, while the plane coupling
along $c$ direction is different inside $J_{\bot}>0$ and between
$J_{\bot}^{'}>0$ blocks, respectively (see, Fig. \ref{FigModel}).
The phase Hamiltonian of the system can now be written:

\begin{eqnarray}
 & H & =-J_{\|}\sum_{\ell}\sum_{i<j}\cos\left[\varphi_{\ell}\left(\mathbf{r}_{i}\right)-\varphi_{\ell}\left(\mathbf{r}_{j}\right)\right]\nonumber \\
 &  & -J_{\bot}\sum_{i}\sum_{m}\sum_{\alpha\ge0}^{n-2}\cos\left[\varphi_{mn+\alpha}\left(\mathbf{r}_{i}\right)-\varphi_{mn+\alpha+1}\left(\mathbf{r}_{i}\right)\right]\nonumber \\
 &  & -J_{\bot}^{'}\sum_{i}\sum_{m}\cos\left[\varphi_{mn+n-1}\left(\mathbf{r}_{i}\right)-\varphi_{mn+n}\left(\mathbf{r}_{i}\right)\right],\label{Hamiltonian}\end{eqnarray}
where indices $i$ and $j$ run over sites in the $ab$-planes of
the $\ell$-th layer, while $\alpha$ numbers the layers within the
$m$-th block. Introducing two-dimensional vectors from Eq. (\ref{SVectors}),
the Hamiltonian reads:\begin{eqnarray}
H & = & -J_{\|}\sum_{\ell}\sum_{i<j}\mathbf{S}_{\ell}\left(\mathbf{r}_{i}\right)\mathbf{S}_{\ell}\left(\mathbf{r}_{j}\right)\nonumber \\
 &  & -J_{\bot}\sum_{i}\sum_{m}\sum_{\alpha\ge0}^{n-2}\mathbf{S}_{mn+\alpha}\left(\mathbf{r}_{i}\right)\mathbf{S}_{mn+\alpha+1}\left(\mathbf{r}_{i}\right)\nonumber \\
 &  & -J_{\bot}^{'}\sum_{i}\sum_{m}\mathbf{S}_{mn+n-1}\left(\mathbf{r}_{i}\right)\mathbf{S}_{mn+n}\left(\mathbf{r}_{i}\right).\label{eq:MultiHamiltonianS}\end{eqnarray}

Furthermore, we make use of the spherical closure relation in Eq.
(\ref{SphericalConstraint}). However, the direct implementation of
the Eq. (\ref{SphericalConstraint}) is obstructed by the lack of
complete translational symmetry: grouping of layers within blocks
breaks translational symmetry along $c$ axis, since the inter-plane
coupling varies with period of $n$, when moving from one plane to
another. As a result, the three-dimensional Fourier transform of the
$\mathbf{S}_{\ell}\left(\mathbf{r}_{i}\right)$ variables in Eq. (\ref{Spherical_Fourier}),
used to diagonalize single-layer Hamiltonian in the previous Section,
cannot be performed. To overcome this difficulty, we implement a combination
of two-dimensional Fourier transform for in-plane vector variables
\begin{equation}
\mathbf{S}_{\ell}\left(\mathbf{r}_{i}\right)=\frac{1}{N_{\|}}\sum_{\mathbf{k}}\mathbf{S}_{\mathbf{k}\ell}e^{-i\mathbf{kr}_{i}}.\end{equation}
and transfer matrix method for one-dimensional decorated structure
along $c$-axis.\cite{DecoratedLattice} This operation diagonalizes
all terms in the Hamiltonian in Eq. (\ref{eq:MultiHamiltonianS})
with respect to $\mathbf{k}$, leaving the dependence on layer index
$\ell$ unchanged. Therefore, the partition function can be written
in the form:\begin{eqnarray}
Z & = & \int_{-\infty}^{+\infty}\frac{d\zeta}{2\pi i}\exp\left\{ N\zeta+\frac{1}{2}\ln\int_{-\infty}^{+\infty}\prod_{\mathbf{k},\ell}d^{2}\mathbf{S}_{\mathbf{k}\ell}\right.\nonumber \\
 &  & \times\left.\exp\left[-\frac{1}{N_{\|}}\sum_{\mathbf{k},\ell,\ell'}\mathbf{S}_{\mathbf{k}\ell}A_{N_{\bot}}^{\ell\ell'}\left(\mathbf{k}\right)\mathbf{S}_{\mathbf{k}\ell'}\right]\right\} ,\label{eq:MultiPartAn}\end{eqnarray}
where $A_{N_{\bot}}^{\ell\ell'}\left(\mathbf{k}\right)$ is an element
of a square $N_{\bot}\times N_{\bot}$ band matrix, appearing as a
result of non-trivial coupling structure along $c$-direction:\begin{widetext}\begin{center}

\begin{equation} \begin{array}{cc} &\hspace{1em}\overbrace{\hspace{12.5em}}^{\begin{array}{c}n\end{array}} \hfill\\ A_{N_\bot }\left( \mathbf{k} \right) = & \left. \left[ \begin{tabular}{ccccccccc} \cline{1-3}\multicolumn{1}{|c}{$\zeta -\frac{\beta J_{\|}\left(\mathbf{k}\right) }{2}$} & $-\frac{\beta J_\bot }{2}$ &\multicolumn{1}{c|}{$\ddots$}&$0$&$0$&$\cdots$&$0$&$0$&$0$ \\\multicolumn{1}{|c}{$-\frac{\beta J_\bot }{2}$}&$\ddots$&\multicolumn{1}{c|}{$-\frac{\beta J_\bot }{2}$}&$0$ &$0$&$\cdots$&$0$&$0$&$0$ \\\multicolumn{1}{|c}{$\ddots$}&$-\frac{\beta J_\bot }{2}$&\multicolumn{1}{c|}{$\zeta -\frac{\beta J_{\|}\left(\mathbf{k}\right) }{2} $}&\fcolorbox[rgb]{0.9,0.9,0.9}{0.9,0.9,0.9}{$-\frac{\beta J_{\bot}^{'}}{2}$}&$0$&$\cdots$&$0$&$0$&$0$ \\\cline{1-5}$0$&$0$&$\overset{\vspace{0.1em}}{\fcolorbox[rgb]{0.9,0.9,0.9}{0.9,0.9,0.9}{$-\frac{\beta J_{\bot}^{'}}{2}$}}$&\multicolumn{1}{|c}{$\zeta -\frac{\beta J_{\|}\left(\mathbf{k}\right) }{2}$}&$-\frac{\beta J_\bot }{2}$&$\ddots$&$\vdots$&$\vdots$&$\vdots$\\$0$&$0$&$0$&\multicolumn{1}{|c}{$-\frac{\beta J_\bot }{2}$}&$\ddots$&\multicolumn{1}{c|}{$-\frac{\beta J_\bot }{2}$}&$0$&$0$&$0$\\$\vdots$&$\vdots$&$\vdots$&$\ddots$&$-\frac{\beta J_\bot }{2}$&\multicolumn{1}{c|}{$\zeta - \frac{\beta J_{\|}\left(\mathbf{k}\right) }{2}$}&\fcolorbox[rgb]{0.9,0.9,0.9}{0.9,0.9,0.9}{$-\frac{\beta J_{\bot}^{'}}{2}$}&$0$&$0$\\\cline{5-9}$0$&$0$&$0$&$\cdots$&$0$&$\overset{\vspace{0.1em}}{\fcolorbox[rgb]{0.9,0.9,0.9}{0.9,0.9,0.9}{$-\frac{\beta J_{\bot}^{'}}{2}$}}$&\multicolumn{1}{|c}{$\zeta -\frac{\beta J_{\|}\left(\mathbf{k}\right) }{2}$} & $-\frac{\beta J_\bot }{2}$ &\multicolumn{1}{c|}{$\ddots$}\\$0$&$0$&$0$&$\cdots$&$0$&$0$&\multicolumn{1}{|c}{$-\frac{\beta J_\bot }{2}$}&$\ddots$&\multicolumn{1}{c|}{$-\frac{\beta J_\bot }{2}$}\\$0$&$0$&$0$&$\cdots$&$0$&$0$&\multicolumn{1}{|c}{$\ddots$}&$-\frac{\beta J_\bot }{2}$&\multicolumn{1}{c|}{$\underset{\vspace{0.1em}}{\zeta -\frac{\beta J_{\|}\left(\mathbf{k}\right) }{2}}$} \\\cline{7-9}\end{tabular} \right] \right\}N_\bot .\\ &\hspace{0.5em}\underbrace{\hfill \hspace{1em} \hspace{37.5em}}_{\begin{array}{c} N_\bot \end{array}} \hfill \end{array} \label{MatrixAN} \end{equation}

\end{center}\end{widetext}where\begin{equation}
J_{\|}\left(\mathbf{k}\right)=2J_{\|}\left[\cos\left(ak_{x}\right)+\cos\left(ak_{y}\right)\right].\end{equation}
Here, $A_{N_{\bot}}$ has square $n\times n$ diagonal blocks:\begin{equation}
B_{n}=\left[\begin{array}{cccc}
\zeta-\frac{\beta J_{\|}\left(\mathbf{k}\right)}{2} & -\frac{\beta J_{\bot}}{2} & 0 & 0\\
-\frac{\beta J_{\bot}}{2} & \ddots & \ddots & 0\\
0 & \ddots & \ddots & -\frac{\beta J_{\bot}}{2}\\
0 & 0 & -\frac{\beta J_{\bot}}{2} & \zeta-\frac{\beta J_{\|}\left(\mathbf{k}\right)}{2}\end{array}\right]\label{MacierzA}\end{equation}
containing interactions inside the stack of $n$ planes. Performing
Gaussian integration in Eq. (\ref{eq:MultiPartAn}), one arrives at
the equation involving determinant of the matrix in Eq. (\ref{MatrixAN}):\begin{equation}
Z=\int_{-\infty}^{+\infty}\frac{d\zeta}{2\pi i}\exp\left[N\zeta+\sum_{\mathbf{k}}\ln\left(\frac{N_{\|}^{\frac{1}{2}N_{\bot}}}{\det A_{N_{\bot}}}\right)\right].\label{PartFunction2}\end{equation}
Thus, the problem of finding the partition function in Eq. (\ref{PartFunction2})
is to find a way to calculate the determinant of the matrix $A_{N_{\bot}}$:\begin{equation}
\det\left(A_{N_{\bot}}\right)\equiv\left|A_{N_{\bot}}\right|.\label{eq:DetAN}\end{equation}
However, because of the band structure of the matrix $A_{N_{\bot}}$
it can be conveniently done using generalised Laplace method. As a
result, the determinant in Eq. (\ref{eq:DetAN}) can be written in
a recurrence form:\begin{equation}
\left|A_{N_{\bot}}\right|=\left|B_{n}\right|\left|A_{N_{\bot}-n}\right|-\left(\frac{\beta J_{\bot}^{^{\prime}}}{2}\right)^{2}\left|B_{n-1}\right|\left|A_{N_{\bot}-n-1}\right|,\label{RowBL}\end{equation}
 where $A_{N_{\bot}-n}$ and $A_{N_{\bot}-n-1}$ arise from removing
the first $n$ and $n+1$ rows and columns of $A_{N_{\bot}}$, respectively.
Similarly to Eq. (\ref{RowBL}), for $\left|A_{N_{\bot}-1}\right|$
the recurrence formula can be written as:\begin{eqnarray}
\left|A_{N_{\bot}-1}\right| & = & \left|B_{n-1}\right|\left|A_{N_{\bot}-n}\right|\nonumber \\
 & - & \left(\frac{\beta J_{\bot}^{^{\prime}}}{2}\right)^{2}\left|B_{n-2}\right|\left|A_{N_{\bot}-n-1}\right|.\label{RowBLm1}\end{eqnarray}
 Defining $\left|B_{0}\right|=1$, the formula in Eq. (\ref{RowBLm1})
holds for any $n\ge2$. The results for the single-layer system ($n=1$)
from the previous Section can be recovered by performing calculations
for any $n\ge2$ and setting $J_{\bot}=J_{\bot}^{'}$. Using Eqs.
(\ref{RowBL}) and (\ref{RowBLm1}) we can write the recurrence relation
for the $A_{N_{\bot}}$ in Eq. (\ref{MatrixAN}) as an operator equation:\begin{equation}
\left[\begin{array}{c}
\left|A_{N_{\bot}}\right|\\
\left|A_{N_{\bot}-1}\right|\end{array}\right]=\mathbf{T}_{1}\left[\begin{array}{c}
\left|A_{N_{\bot}-n}\right|\\
\left|A_{N_{\bot}-n-1}\right|\end{array}\right],\label{Transfer1}\end{equation}
where $\mathbf{T}_{1}$ is transfer matrix:\begin{equation}
\mathbf{T}_{1}=\left[\begin{array}{ll}
\left|B_{n}\right| & -\left({\displaystyle \frac{\beta J_{\bot}^{'}}{2}}\right)^{2}\left|B_{n-1}\right|\\
\left|B_{n-1}\right| & -\left({\displaystyle \frac{\beta J_{\bot}^{'}}{2}}\right)^{2}\left|B_{n-2}\right|\end{array}\right].\end{equation}
Applying the rule in Eq. (\ref{Transfer1}) recursively, one obtains:\begin{equation}
\left[\begin{array}{c}
\left|A_{N_{\bot}}\right|\\
\left|A_{N_{\bot}-1}\right|\end{array}\right]=\mathbf{T}_{1}^{\frac{1}{n}N_{\bot}-1}\left[\begin{array}{c}
\left|A_{n}\right|\\
\left|A_{n-1}\right|\end{array}\right].\label{eq:TransEq1}\end{equation}
Therefore, finding the solution of the Eq. (\ref{eq:TransEq1}) reduces
to the problem of determining the power of the transfer matrix operator
$\mathbf{T}_{1}.$ To this end we diagonalize the matrix $\mathbf{T}_{1}$
with the help of a unitary transformation $\mathcal{U}_{1}$ (see,
Appendix \ref{sec:Matrix-diagonalization}):\begin{equation}
\mathbf{T}_{1}=\mathcal{U}_{1}\left[\begin{array}{cc}
\lambda_{1} & 0\\
0 & \lambda_{2}\end{array}\right]\mathcal{U}_{1}^{-1},\label{eq:TransMatrixT1}\end{equation}
where $\lambda_{1}$ and $\lambda_{2}$ are eigenvalues of $\mathbf{T}_{1}$,
whereas $\mathcal{U}_{1}$ is a unitary matrix built of eigenvectors
of $\mathbf{T}_{1}$. Therefore,\begin{equation}
\mathbf{T}_{1}^{\frac{1}{n}N_{\bot}-1}=\mathcal{U}_{1}\left[\begin{array}{cc}
\lambda_{1}^{\frac{1}{n}N_{\bot}-1} & 0\\
0 & \lambda_{2}^{\frac{1}{n}N_{\bot}-1}\end{array}\right]\mathcal{U}_{1}^{-1}.\label{eq:TransMatrixT1Power}\end{equation}
The eigenvalue $\lambda_{1}$ is chosen to satisfy $\lambda_{1}>\lambda_{2}$.
In the large-$N_{\bot}$ (thermodynamic) limit only the largest eigenvalue
$\lambda_{1}$ will contribute to the statistical sum in Eq. (\ref{PartFunction2}),

\begin{eqnarray}
\left[\begin{array}{cc}
\lambda_{1} & 0\\
0 & \lambda_{2}\end{array}\right]^{\frac{1}{n}N_{\bot}-1} & = & \left\{ \lambda_{1}\left[\begin{array}{cc}
1 & 0\\
0 & \lambda_{2}/\lambda_{1}\end{array}\right]\right\} ^{\frac{1}{n}N_{\bot}-1}\nonumber \\
 & \underset{N_{\bot}\rightarrow\infty}{=} & \lambda_{1}^{\frac{1}{n}N_{\bot}-1}\left[\begin{array}{cc}
1 & 0\\
0 & 0\end{array}\right].\label{eq:L1LargeN}\end{eqnarray}
 Combining Eqs. (\ref{eq:TransEq1}), (\ref{eq:TransMatrixT1Power})
and (\ref{eq:L1LargeN}) we obtain finally:\begin{equation}
\left[\begin{array}{c}
\left|A_{N_{\bot}}\right|\\
\left|A_{N_{\bot}-1}\right|\end{array}\right]\underset{N_{\bot}\rightarrow\infty}{=}\lambda_{1}^{\frac{1}{n}N_{\bot}-1}\mathcal{U}_{1}\left[\begin{array}{cc}
1 & 0\\
0 & 0\end{array}\right]\mathcal{U}_{1}^{-1}\left[\begin{array}{c}
\left|A_{n}\right|\\
\left|A_{n-1}\right|\end{array}\right],\label{RowBLFinal}\end{equation}
where:\begin{widetext}\begin{equation}
\lambda_{1}=\frac{1}{2}\left\{ \left|B_{n}\right|-\left(\frac{\beta J_{\bot}^{^{\prime}}}{2}\right)^{2}\left|B_{n-2}\right|+\sqrt{\left[\left|B_{n}\right|+\left(\frac{\beta J_{\bot}^{^{\prime}}}{2}\right)^{2}\left|B_{n-2}\right|\right]^{2}-\left(\beta J_{\bot}^{^{\prime}}\right)^{2}\left|B_{n-1}\right|^{2}}\right\} .\label{eq:DefLambda1}\end{equation}
 From Eq. (\ref{RowBLFinal}) one can extract the value of determinant
of the matrix $A_{N_{\bot}}$:\begin{equation}
\left|A_{N_{\bot}}\right|=\lambda_{1}^{\frac{1}{n}N_{\bot}-1}\left(\mathcal{U}\right)_{11}\left[\left(\mathcal{U}_{1}^{-1}\right)_{11}\left|B_{n}\right|+\left(\mathcal{U}_{1}^{-1}\right)_{21}\left|B_{n-1}\right|\right],\end{equation}
where $\left(\mathcal{U}\right)_{\mu\nu}$ and $\left(\mathcal{U}_{1}^{-1}\right)_{\mu\nu}$
denote elements of the matrix $\mathcal{U}_{1}$ and the inverse matrix
$\mathcal{U}_{1}^{-1}$, respectively (see, Appendix \ref{sec:Matrix-diagonalization}).
Explicitly,\begin{align}
\left|A_{N_{\bot}}\right| & =\mathcal{C}\left[\frac{\left|B_{n}\right|-\left({\displaystyle \frac{\beta J_{\bot}^{^{\prime}}}{2}}\right)^{2}\left|B_{n-2}\right|}{2}+\frac{\sqrt{\left[\left|B_{n}\right|+\left({\displaystyle \frac{\beta J_{\bot}^{^{\prime}}}{2}}\right)^{2}\left|B_{n-2}\right|\right]^{2}-\left({\displaystyle \beta J_{\bot}^{^{\prime}}}\right)^{2}\left|B_{n-1}\right|^{2}}}{2}\right]^{\frac{1}{n}N_{\bot}-1},\label{eq:DetANFinal}\end{align}
 \end{widetext}is the function of determinants of matrices $B_{n}$
with intra-block interactions $J_{\bot}$ and \begin{equation}
\mathcal{C}=\left(\mathcal{U}_{1}\right)_{11}\left[\left(\mathcal{U}_{1}^{-1}\right)_{11}\left|B_{n}\right|+\left(\mathcal{U}_{1}^{-1}\right)_{21}\left|B_{n-1}\right|\right].\label{eq:DefC}\end{equation}

The determinant of the matrix $B_{n}$ in Eq. (\ref{MacierzA}) can
be found using similar method by employing the Laplace expansion (see
Eq. (\ref{RowBL})-(\ref{RowBLFinal})). However, since the order
of $B_{n}$ is finite, both eigenvalues of the new transfer matrix
$\mathbf{T}_{2}$ must be taken into account:\begin{eqnarray}
 &  & \left[\begin{array}{c}
\left|B_{n+1}\right|\\
\left|B_{n}\right|\end{array}\right]=\mathbf{T}_{2}^{n/2-1}\left[\begin{array}{c}
\left|B_{2}\right|\\
\left|B_{1}\right|\end{array}\right]=\mathcal{U}_{2}\nonumber \\
 &  & \,\,\,\,\,\,\,\,\,\,\,\,\,\,\,\,\,\,\,\,\times\left[\begin{array}{cc}
\lambda_{+}^{n/2-1} & 0\\
0 & \lambda_{-}^{n/2-1}\end{array}\right]\mathcal{U}_{2}^{-1}\left[\begin{array}{c}
\left|B_{2}\right|\\
\left|B_{1}\right|\end{array}\right],\label{eq:TransMatrixT2}\end{eqnarray}
where transfer matrix $\mathbf{T}_{2}$ is:\begin{equation}
\mathbf{T}_{2}=\left[\begin{array}{ll}
\left[\zeta-\frac{\beta J_{\|}\left(\mathbf{k}\right)}{2}\right]^{2}-\frac{\beta^{2}J_{\bot}^{2}}{4}\,\, & -\frac{\beta^{2}J_{\bot}^{2}}{4}\left[\zeta-\frac{\beta J_{\|}\left(\mathbf{k}\right)}{2}\right]\\
\zeta-\frac{\beta J_{\|}\left(\mathbf{k}\right)}{2} & -\frac{\beta^{2}J_{\bot}^{2}}{4}\end{array}\right],\end{equation}
$\mathcal{U}_{2}$ is matrix built of eigenvectors of $\mathbf{T}_{2}$
and eigenvalues $\lambda_{+},\lambda_{-}$ of $\mathbf{T}_{2}$and
read:\begin{eqnarray}
\lambda_{\pm} & = & \frac{1}{2}\left\{ \left[\zeta-\frac{\beta J_{\|}\left(\mathbf{k}\right)}{2}\right]^{2}-\frac{\beta^{2}J_{\bot}^{2}}{2}\pm\left[\zeta-\frac{\beta J_{\|}\left(\mathbf{k}\right)}{2}\right]\right.\nonumber \\
 & \times & \left.\sqrt{\left[\zeta-\frac{\beta J_{\|}\left(\mathbf{k}\right)}{2}\right]^{2}-\left(\beta J_{\bot}\right)^{2}}\right\} .\label{eq:LambdaPlusMinus}\end{eqnarray}
Finally, we arrive at the value of $\left|B_{n}\right|$ determinant:\begin{align}
\left|B_{n}\right| & =\frac{\left|B_{2}\right|}{2}\left(\lambda_{+}^{\frac{n}{2}}+\lambda_{-}^{\frac{n}{2}}\right)+\frac{\left|B_{1}\right|}{2}\left(\lambda_{+}^{\frac{n}{2}}-\lambda_{-}^{\frac{n}{2}}\right)\nonumber \\
 & \times\frac{2\left|B_{2}\right|-\left(\beta J_{\bot}\right)^{2}}{\sqrt{\left[\zeta-{\displaystyle \frac{\beta J_{\|}\left(\mathbf{k}\right)}{2}}\right]^{2}-\left({\displaystyle \beta J_{\bot}}\right)^{2}}}.\label{eq:DetBNFinal}\end{align}
with $\lambda_{+}$ and $\lambda_{-}$ given by Eq. (\ref{eq:LambdaPlusMinus}).
Thus, with the result in Eq. (\ref{eq:DetBNFinal}), the calculation
of the determinant of the matrix $A_{N_{\bot}}$ in Eq. (\ref{eq:DetANFinal})
and thereby the partition function given by Eq. (\ref{PartFunction2})
is completed.

\section{Critical temperature}

Using the partition function in Eq. (\ref{PartFunction2}) with the
definitions in Eqs. (\ref{eq:DefLambda1}) and (\ref{eq:DefC}) the
free energy $f$ is given by:\begin{eqnarray}
f & = & \frac{\zeta}{\beta}-\frac{1}{\beta N_{\bot}N_{\|}}\sum_{\mathbf{k}}\ln\lambda_{1}^{\frac{1}{n}N_{\bot}-1}\left(\mathbf{k}\right)+\frac{1}{2\beta N_{\|}}\ln N_{\|}\nonumber \\
 &  & -\frac{1}{\beta N_{\bot}N_{\|}}\sum_{\mathbf{k}}\ln\mathcal{C}\nonumber \\
 &  & \underset{N_{\|}N_{\bot}\rightarrow\infty}{\longrightarrow}\frac{\zeta}{\beta}-\frac{1}{\beta n}\int_{-\frac{\pi}{a}}^{\frac{\pi}{a}}\frac{d^{2}\mathbf{k}}{\left(2\pi/a\right)^{2}}\ln\lambda_{1}\left(\mathbf{k}\right)\end{eqnarray}
 while the saddle-point equation for the spherical Lagrange multiplier
becomes:

\begin{equation}
\left.\frac{\partial f\left(\zeta\right)}{\partial\zeta}\right|_{\zeta=\zeta_{0}}=0.\end{equation}
Explicitly,\begin{equation}
1=\frac{1}{n}\int_{-\frac{\pi}{a}}^{\frac{\pi}{a}}\frac{d^{2}\mathbf{k}}{\left(2\pi/a\right)^{2}}\left.\frac{\frac{\partial C_{n}\left(\mathbf{k}\right)}{\partial\zeta}\sqrt{D_{n}\left(\mathbf{k}\right)}+\frac{1}{2}\frac{\partial D_{n}\left(\mathbf{k}\right)}{\partial\zeta}}{C_{n}\left(\mathbf{k}\right)\sqrt{D_{n}\left(\mathbf{k}\right)}+D_{n}\left(\mathbf{k}\right)}\right|_{\zeta=\zeta_{0}},\label{SaddlePointFinal}\end{equation}
where we have defined:\begin{align}
C_{n}\left(\mathbf{k}\right) & =\left|B_{n}\right|-\left(\frac{\beta J_{\bot}^{^{\prime}}}{2}\right)^{2}\left|B_{n-2}\right|\nonumber \\
D_{n}\left(\mathbf{k}\right) & =\left[\left|B_{n}\right|+\left(\frac{\beta J_{\bot}^{^{\prime}}}{2}\right)^{2}\left|B_{n-2}\right|\right]^{2}\nonumber \\
 & -\left(\beta J_{\bot}^{^{\prime}}\right)^{2}\left|B_{n-1}\right|^{2}.\end{align}
The in-plane momentum dependence of $\lambda_{1}\left(\mathbf{k}\right)$,
$C_{n}\left(\mathbf{k}\right)$ and $D_{n}\left(\mathbf{k}\right)$
comes from $J_{||}\left(\mathbf{k}\right)$ dependence of $\left|B_{n}\right|$,
see Eq. (\ref{eq:DetBNFinal}). The saddle-point value $\zeta_{0}$
can be determined from the condition that in criticality order-parameter
susceptibility becomes infinite (see, Eq. (\ref{SphericalCorrelator})):\begin{equation}
\left.\frac{C_{n}\left(\mathbf{k}\right)\sqrt{D_{n}\left(\mathbf{k}\right)}+D_{n}\left(\mathbf{k}\right)}{{\displaystyle \frac{\partial C_{n}\left(\mathbf{k}\right)}{\partial\zeta}}\sqrt{D_{n}\left(\mathbf{k}\right)}+\frac{1}{2}{\displaystyle \frac{\partial D_{n}\left(\mathbf{k}\right)}{\partial\zeta}}}\right|_{\begin{array}{l}
\mathbf{k}=0\\
\zeta=\zeta_{0}\end{array}}=0.\label{eq:MultiSpherLagrange}\end{equation}
Closed formulas for the critical temperature can be easily obtained
only for small values of $n$. For example, in the case of a bilayer
system ($n=2$), the respective determinants read:\begin{align}
\left|B_{2}\right| & =\left[\zeta-\frac{\beta J_{\|}\left(\mathbf{k}\right)}{2}\right]^{2}-\left(\frac{\beta J_{\bot}}{2}\right)^{2},\nonumber \\
\left|B_{1}\right| & =\zeta-\frac{\beta J_{\|}\left(\mathbf{k}\right)}{2},\nonumber \\
\left|B_{0}\right| & =1,\end{align}
 and the saddle point equation in Eq. (\ref{SaddlePointFinal}) can
be written explicitly:\begin{eqnarray}
1 & = & \int_{-\frac{\pi}{a}}^{\frac{\pi}{a}}\frac{d^{2}\mathbf{k}}{\left(2\pi/a\right)^{2}}\frac{\zeta-\frac{1}{2}\beta J_{\|}\left(\mathbf{k}\right)}{\sqrt{\left[\zeta_{0}-\frac{\beta}{2}\left(J_{\|}\left(\mathbf{k}\right)-J_{\bot}^{^{\prime}}\right)\right]^{2}-\frac{\beta^{2}J_{\bot}^{2}}{4}}}\nonumber \\
 &  & \times\frac{1}{\sqrt{\left[\zeta_{0}-\frac{\beta}{2}\left(J_{\|}\left(\mathbf{k}\right)+J_{\bot}^{^{\prime}}\right)\right]^{2}-\frac{\beta^{2}J_{\bot}^{2}}{4}}}\end{eqnarray}
 with the Lagrange multiplier:\begin{equation}
\zeta_{0}=\frac{\beta}{2}\left[J_{\|}\left(\mathbf{k}=0\right)+J_{\bot}+J_{\bot}^{^{\prime}}\right].\end{equation}
Finally, the critical temperature of this system (cf. Eq. (\ref{SphericalCritTemp}))
is given by\begin{eqnarray}
\beta_{c} & = & \frac{1}{2}\int_{-\infty}^{+\infty}d\xi\rho\left(\xi\right)\frac{2\left(2-\xi\right)J_{\|}+J_{\bot}+J_{\bot}^{^{\prime}}}{\sqrt{\left(2-\xi\right)J_{\|}\left[\left(2-\xi\right)J_{\|}+J_{\bot}\right]}}\nonumber \\
 &  & \times\frac{1}{\sqrt{\left[2\left(2-\xi\right)J_{\|}+\left(J_{\bot}+2J_{\bot}^{^{\prime}}\right)\right]^{2}-J_{\bot}^{2}}}\end{eqnarray}
with\begin{equation}
\rho\left(\xi\right)=\int_{-\frac{\pi}{a}}^{\frac{\pi}{a}}\frac{d^{2}\mathbf{k}}{\left(2\pi/a\right)^{2}}\delta\left[\xi-\cos\left(ak_{x}\right)-\cos\left(ak_{y}\right)\right]\end{equation}
 being density of states of the square lattice given by:\begin{equation}
\rho\left(\xi\right)=\frac{1}{\pi^{2}}\mathbf{K}\left(\sqrt{1-\frac{\xi^{2}}{4}}\right)\,\Theta\left(1-\frac{\left|\xi\right|}{2}\right),\end{equation}
 where $\Theta\left(x\right)$ is the unit-step function. For higher
values of $n$ we can resort to direct numerical evaluation of Eqs.
(\ref{SaddlePointFinal}) and (\ref{eq:MultiSpherLagrange}) in order
to compute $T_{c}\left(n\right)$.

\begin{figure}
\includegraphics[%
  scale=0.35]{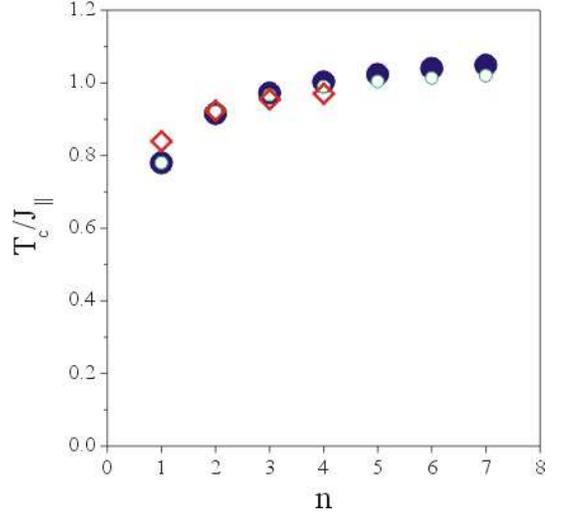}

\caption{(Color online) Critical temperature vs. number of \hbox{Cu-O} layers
within block for $J_{\bot}/J_{\|}=0.1$ and $J_{\bot}^{'}/J_{\|}=0.01$
(blue filled circles). Empty circles (green) denote the critical temperature
dependence $T_{c}\left(n\right)=T_{c}\left(1\right)+0.28\left(1-1/n\right)$,
from Ref. \onlinecite{Leggett} (factor of $0.28$ was chosen to fit
the results of the present paper). Diamonds (red) represent results
from Monte-Carlo (MC) simulations of the classical XY model (Carlson,
et al., Ref. \onlinecite{Carlson}). The data of Ref. \onlinecite{Carlson}
is scaled by overall factor of $0.77$ to accommodate the finite size
effect of MC simulations. \label{Fig1}}
\end{figure}

\begin{figure}
\includegraphics[%
  scale=0.35]{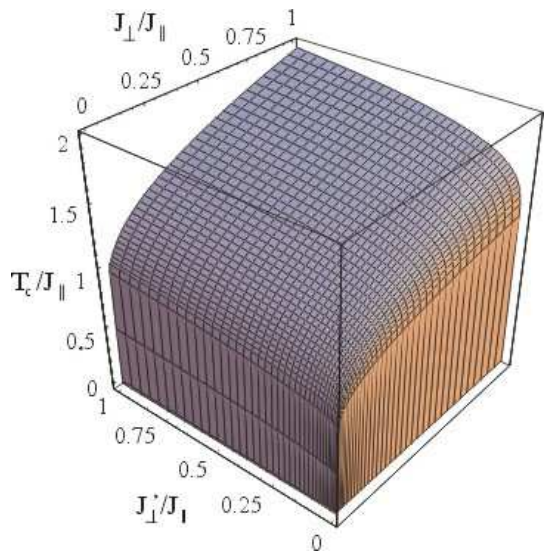}

\caption{(Color online) Critical temperature $T_{c}\left(n\right)$ for $n=4$
as a function of $c$-axis phase stiffnesses $J_{\bot}$ and $J_{\bot}^{'}$.}

\label{Fig3D}
\end{figure}

\begin{figure}
\includegraphics[%
  scale=0.35]{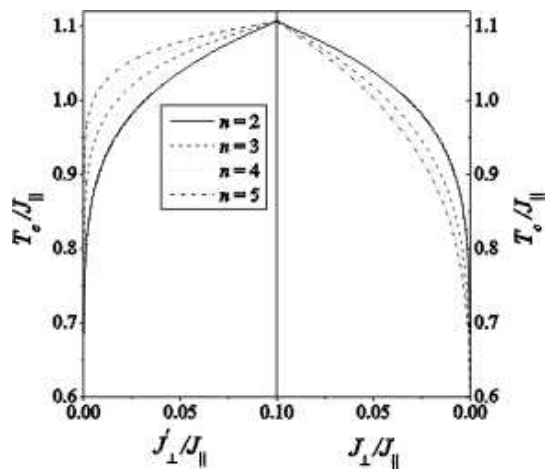}

\caption{Critical temperature $T_{c}\left(n\right)$ as a function of $c$-axis
phase stiffnesses $J_{\bot}$ and $J_{\bot}^{'}$ for various values
of $n$ as indicated in the inset.}

\label{FigJoty}
\end{figure}

\begin{figure}
\includegraphics[%
  scale=0.35]{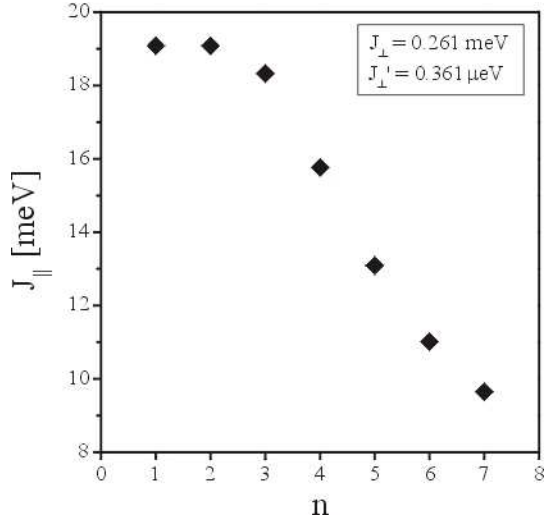}

\caption{In-plane phase stiffness $J_{\|}$, from Table \ref{TabJrown}, as
a function of number of layers $n$. }

\label{Fig_J}
\end{figure}

The dependence of the critical temperature on the number of layers
$n$ in a block was presented in Fig. \ref{Fig1}. Phase stiffnesses
$J_{\|}$, $J_{\bot}$ and $J_{\bot}^{'}$ were chosen to satisfy
\begin{equation}
J_{\|}\gg J_{\bot}>J_{\bot}^{'},\end{equation}
which is physically reasonable, since the interlayer couplings $J_{\bot}$
and $J_{\bot}^{'}$ are much smaller than in-plane phase stiffness
$J_{\|}$ and inter-layer (intra-block) phase stiffness $J_{\bot}$
is greater than inter-block coupling $J_{\bot}^{'}$. We found that
the critical temperature increases monotonically with $n$. This has
a simple explanation: for $n=1$ (single-layer system) with the fixed
phase stiffness $J_{\|}$, the critical temperature is determined
by $J_{\bot}^{'}$, see Fig. \ref{Fig312D}. On the other hand, in
the limit of $n\rightarrow\infty$, $n$-layer blocks are of infinite
size and inter-block coupling $J_{\bot}^{'}$ is effectively replaced
by $J_{\bot}$. Because $J_{\bot}>J_{\bot}^{'}$, according to the
result from the previous Section, $T_{c}\left(n=\infty\right)>T_{c}\left(n=1\right)$.
For intermediate values on $n$, one can expect a monotonic increase
of the $T_{c}\left(n\right)$ with increasing $n$, as depicted in
Fig. \ref{Fig1}. In this respect, our findings are similar to the
results from Ref. \onlinecite{Leggett}, where the increase of the
critical temperature of \emph{identically} doped members of the same
homologous series due to the interlayer Coulomb interaction was found
to be \begin{equation}
T_{c}\left(n\right)=T_{c}\left(1\right)+\mathrm{const}\times\left(1-\frac{1}{n}\right).\end{equation}
Similar results were obtained by Monte-Carlo simulations,\cite{Carlson}
where $T_{c}\left(n\right)$ was computed numerically by means of
the Binder parameter\cite{Binder} for systems of size up to $24\times24\times24$.
Additionally, we present the influence of inter-planar coupling constants
$J_{\bot}$ and $J_{\bot}^{'}$ on $T_{c}$, see Figs. \ref{Fig3D}
and \ref{FigJoty}. For a bilayer system the role of both interaction
constants is identical. For systems with $n\ge3$ the influence of
the inter-block coupling $J_{\bot}^{'}$ is less visible, however
$J_{\bot}^{'}=0$ suppresses the critical temperature to zero, since
for finite $n$ the isolated stack of $n$ coupled layers is effectively
quasi two-dimensional.%
Therefore, we conclude that the change of the number $n$ of layers
within block cannot alone lead to the appearance of the maximum in
the critical temperature $T_{c}$ as a function of $n$ for fixed
values of the microscopic phase stiffnesses. It is necessary to introduce
another factor, which acts competitively to the effective increasing
of inter-layer coupling (for $J_{\bot}>J_{\bot}^{'}$) triggered by
increase of $n$. As in homologous series of superconducting cuprates
in-plane lattice parameters and distances between \hbox{Cu-O} planes
are constant among different members of the family (see, Ref. \onlinecite{HgBCCOLatticeParams}),
one can expect that the change of $n$ will mainly affect $J_{\|}$
coupling parameter. Without assuming any specific mechanism or introducing
a competing (hidden) order, one can calculate values of $J_{\|}$
for given $J_{\bot}$ and $J_{\bot}^{'}$ leading to dependence of
critical temperature on $n$, which is in agreement with experimental
findings. The result is presented in Fig. \ref{Fig_J} and in Table
\ref{TabJrown}. It is visible that for $n=2,\,3$ depicted dependence
changes from rising to falling. This coincides with the fact that
for $n=3$ layers lose their equivalence i.e. they split for two groups:
outer, lying on block surfaces and inner. Because, in real materials,
within block of \hbox{Cu-O} planes $c$-axis redistribution of charge
is possible, one can expect that the nonequivalence of the planes
within a block may lead to renormalization of the in-plane phase stiffnesses
of the respective layers.

\begin{table}
\begin{tabular}{|c|c|c|}
\hline 
$\hspace{0.9cm}n\hspace{0.9cm}$&
$\hspace{0.6cm}T_{c}\,[\mathrm{K}]\hspace{0.6cm}$&
$\hspace{0.5cm}J_{\|}\,[\mathrm{meV}]\hspace{0.5cm}$\tabularnewline
\hline
\hline 
1&
97&
19.1\tabularnewline
\hline 
2&
126&
19.1\tabularnewline
\hline 
3&
135&
18.3\tabularnewline
\hline 
4&
125&
15.8\tabularnewline
\hline 
5&
110&
13.1\tabularnewline
\hline 
6&
97&
11.0\tabularnewline
\hline 
7&
88&
9.64\tabularnewline
\hline
\end{tabular}

\caption{Superconducting critical temperature $T_{c}\left(n\right)$ of \hbox{HgBa$_2$Ca$_{n-1}$Cu$_n$O$_{2n+2+\delta }$}
compounds taken from Ref. \onlinecite{MultiLayer1} and the corresponding
values of the in-plane stiffnesses $J_{\|}$ calculated from Eq. (\ref{SaddlePointFinal})
for $J_{\bot}=0.261\mathrm{meV}$ and $J_{\bot}^{'}=0.000361\mathrm{meV}$.}

\label{TabJrown}
\end{table}

\section{Charge redistribution}

In this section, using the result of the previous paragraphs, we present
a phenomenological theory for the downturn of the $T_{c}\left(n\right)$
with increasing number of layers in multilayer materials and compare
it against the observed systematics, providing support to the presented
theory. The idea is simple: it is well known that a superconductor
is characterized by a global phase stiffness, which is proportional
to the condensate density. In high-$T_{c}$ cuprates the global phase
stiffness is strongly doping dependent and this dependence has a universal
character as evidenced experimentally by the dome-shaped dependence
of the critical temperature as a function of doping. Phrased in the
language of the presented model with phase variables and microscopic
phase stiffnesses, it means that the parameters of this model have
to be changed substantially with doping, since the global phase stiffness
(i.e. the condensate) builds on microscopic phase stiffnesses and
phase variables. The dependence of the phase stiffnesses on doping
$\delta$ obviously cannot be found in the presented model, since
they are treated as effective values to be determined, in principle,
from the (yet unavailable) microscopic theory. However, assuming that
all of them share the same doping dependence:\begin{eqnarray}
J_{\|} & = & g\left(\delta\right)\tilde{J}_{\|}\nonumber \\
J_{\bot} & = & g\left(\delta\right)\tilde{J}_{\bot}\nonumber \\
J_{\bot}^{'} & = & g\left(\delta\right)\tilde{J}_{\bot}^{'},\label{eq:JotTilde}\end{eqnarray}
one can rewrite the saddle point equation in Eq. (\ref{SaddlePointFinal})
in the form:\begin{equation}
T_{c}\left(\delta\right)=g\left(\delta\right)h\left(\tilde{J}_{\|},\tilde{J}_{\bot},\tilde{J}_{\bot}^{'}\right),\label{eq:SaddlePointCharge}\end{equation}
where $h$ is a function that is independent on doping $\delta$.
For a given microscopic phase stiffnesses it follows from Eq. (\ref{eq:SaddlePointCharge})
that the function $g\left(\delta\right)$ determines the overall doping
dependence of $T_{c}\left(\delta\right)$. However, the parabolic
shape of $T_{c}$$\left(\delta\right)$ is well-known (see, Ref. \onlinecite{BSCCOPhaseDiagram})
and the superconducting {}``dome'' in the $T-\delta$ phase diagram
extends from $\delta\approx0.05$ to $\delta\approx0.25$ with the
optimal doping $\delta_{opt}\approx0.15$. This allows us to extract
the doping dependence of $g\left(\delta\right)$ as a quadratic function
of $\delta$ in the form:\begin{equation}
g\left(\delta\right)=\left\{ \begin{array}{l}
1-\frac{1}{\alpha}\left(\delta-\delta_{opt}\right)^{2}\,\,\mathrm{for}\,0.05<\delta<0.025\\
0\,\,\,\mathrm{for}\,\delta<0.05\,\mathrm{or\,\delta>0.25},\end{array}\right.\end{equation}
where $\alpha=0.01$. According to results of the previous section,
in the case of strong anisotropy ($J_{\|}\gg J_{\bot},J_{\bot}^{'}$),
the in-plane phase stiffness $J_{\|}$ has the major influence on
the critical temperature $T_{c}$, which is weakly dependent on inter-planar
phase stiffnesses (see, Fig. \ref{Fig312D}). Therefore, we limit
further discussion to the case, where the inter-planar phase stiffnesses
as functions of doping $J_{\bot}\approx\tilde{J}_{\bot}$ and $J_{\bot}^{'}\approx\tilde{J}_{\bot}^{'}$
are approximately constant. The value of $\tilde{J}_{\|}$ can be
determined using the following procedure. Obviously, single- and double-layer
systems cannot exhibit charge redistribution: all the planes are equivalent
and therefore, the layers can be optimally doped $g\left(\delta_{opt}\right)=1$.
Using the data for the critical temperature $T_{c}\left(n\right)$
for optimally-doped mercuro-cuprates\cite{MultiLayer1} for $n=1,\,2$,
we can determine $\tilde{J}_{\|}$ and the ratio $\tilde{J}_{\bot}^{'}/\tilde{J}_{\bot}$.
Unfortunately, the estimates of the values of couplings along $c$-axis
$\tilde{J}_{\bot}$ and $\tilde{J}_{\bot}^{'}$ are difficult to make.
However, using the relation between zero-temperature $c$-axis and
$ab$ penetration depths and bare phase stiffnesses:\cite{Roddick}\begin{eqnarray}
\frac{1}{\lambda_{\|}^{2}\left(T=0\right)} & \sim & \tilde{J}_{\|}\nonumber \\
\frac{1}{\lambda_{\bot}^{2}\left(T=0\right)} & \sim & \tilde{J}_{\bot}^{'}\label{eq:LambdaAndPhaseStiff}\end{eqnarray}
we can evaluate $\tilde{J}_{\bot}^{'}/\tilde{J}_{\|}$ to determine
the value of the weakest coupling $\tilde{J}_{\bot}^{'}$. In order
to do this, we assume the largest anisotropy ratio \begin{equation}
\lambda_{\bot}/\lambda_{\|}\sim230\label{eq:LambdaRatio}\end{equation}
 found in cuprates.\cite{PenDepth} In this way all the phase stiffnesses
can be fixed; we found that $\tilde{J}_{\|}=19.1\mathrm{meV}$ (see,
Table \ref{TabJrown}), while $\tilde{J}_{\bot}=0.26\mathrm{meV}$
and $\tilde{J}_{\bot}^{'}=0.00036\mathrm{meV}$. Although, the estimates
of $c$-axis phase stiffnesses in general can substantially vary their
influence on the critical temperature is rather weak, as shown in
Section II. 

\begin{figure}
\includegraphics[%
  scale=0.4]{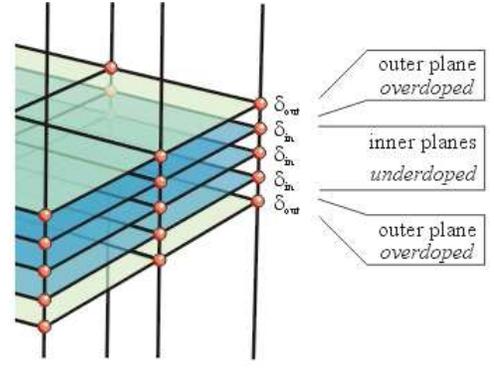}

\caption{(Color online) Schematic representation of charge redistribution
leading to overdoping of outer and underdoping of inner layers, respectively.\label{fig:Charge-redistribution}}
\end{figure}

In the multi-layer systems ($n\ge3$) the planes are not equivalent
any more and charge redistribution within blocks is possible.\cite{ChargeDistrib,ChargeDistrib2}
Therefore, in-plane phase stiffness of every layer $\ell$ becomes
$J_{\|}\left(\delta_{\ell}\right)$. Since our model does not contain
independent values of $J_{\|}$ for different planes (which would
complicate theoretical considerations considerably -- given the lack
of the translational symmetry along $c$-axis), we modify Eq. (\ref{eq:JotTilde})
and use the average over $n$ layers in the $n$-layer block:\begin{equation}
J_{\|}\rightarrow J_{\|}^{av}\left(\delta\right)=\frac{\tilde{J}_{\|}}{n}\sum_{i=1}^{n}g\left(\delta_{i}\right).\label{eq:JAve}\end{equation}
 However, nuclear magnetic resonance (NMR) measurements show that
charge distribution among inner planes is almost constant.\cite{ChargeDistrib}
Therefore, we characterize charge distribution by two parameters:
we assume that the doping of inner and outer planes is given by $\delta_{ip}$
and $\delta_{op}$, respectively, see Fig. \ref{fig:Charge-redistribution}.
Introducing a ratio of dopings for outer and inner planes \begin{equation}
R=\delta_{op}/\delta_{ip}>1,\end{equation}
and noting that in every block there are 2 outer and $n-2$ inner
layers, we can write Eq. (\ref{eq:JAve}) as:\begin{equation}
J_{\|}^{av}\left(\delta_{ip},R,n\right)=\frac{\tilde{J}_{\|}}{n}\left[2g\left(R\delta_{ip}\right)+\left(n-2\right)g\left(\delta_{ip}\right)\right].\label{eq:JAve2}\end{equation}

By observing that the maximum of the critical temperature $T_{c}\left(n\right)$
corresponds to the maximum of $J_{\|}^{av}\left(\delta_{ip},R,n\right)$,
we can find the optimal value of $\delta_{ip}=\delta_{opt}^{*}$ using
the condition\begin{equation}
\left.\frac{dJ_{\|}^{av}\left(\delta_{ip},R,n\right)}{d\delta_{ip}}\right|_{\delta_{ip}=\delta_{opt}^{*}}=0.\end{equation}
Explicitly, using Eq. (\ref{eq:JAve2}) we obtain for $\delta_{opt}^{*}$:\begin{equation}
\delta_{opt}^{*}=\delta_{opt}^{2}\frac{n-2+2R}{n-2+2R^{2}},\end{equation}
which allows to find the value of the \emph{effective} in-plane phase
stiffness \begin{equation}
J_{\|}\equiv J_{\|}^{av}\left(n,R\right)\end{equation}
 as a function of number of layers $n$ and charge imbalance ratio
$R$:\begin{equation}
J_{\|}^{av}\left(n,R\right)=\tilde{J}_{\|}\left[1-\frac{\delta_{opt}^{2}\left(n-2\right)\left(R-1\right)^{2}}{\alpha n\left(n-2+2R^{2}\right)}\right].\label{eq:JAveFinal}\end{equation}

Using Eq. (\ref{eq:JAveFinal}) we can determine the charge imbalance
parameters $R$ as a function of number of layers $n$ for a given
set of experimental data for the critical temperature $T_{c}\left(n\right)$.
For this purpose we use the dependence of the superconducting critical
temperature on number of layers reported by Kuzemskaya, et al., Ref.
\onlinecite{MultiLayer1}. The procedure reads as follows: 

(i) For $n=1$ and $n=2$ by definition $R=1$ and consequently $J_{\|}^{av}\left(n,R\right)\equiv\tilde{J}_{\|}$.
Using the value of $T_{c}\left(1\right)$ from experiment, we can
determine the in-plane $\tilde{J}_{\|}$ and $\tilde{J}_{\bot}^{'}\equiv\tilde{J}_{\bot}$
(for $n=1$) phase stiffnesses using Eq. (\ref{SaddlePointFinal})
for the critical temperature $T_{c}\left(n=1\right)$ and the supplementary
equations (\ref{eq:LambdaAndPhaseStiff}) and (\ref{eq:LambdaRatio}).

(ii) For $n=2$ we determine in-block $\tilde{J}_{\bot}$ phase stiffness
using Eq. (\ref{SaddlePointFinal}) for $T_{c}\left(n=2\right)$,
while keeping $\tilde{J}_{\|}$ and $\tilde{J}_{\bot}^{'}$ unchanged.
In this way all the energy setting parameters $\tilde{J}_{\|}$, $\tilde{J}_{\bot}$
and $\tilde{J}_{\bot}^{'}$ are fixed. 

(iii) For $n>2$, with the computed parameters $\tilde{J}_{\|}$,
$\tilde{J}_{\bot}$ and $\tilde{J}_{\bot}^{'}$, by making use of
Eq. (\ref{SaddlePointFinal}) for theoretical $T_{c}\left(n\right)$,
we calculate the values of $R$, using $T_{c}\left(n\right)$ as an
input from experiment.

\begin{figure}
\includegraphics[%
  scale=0.35]{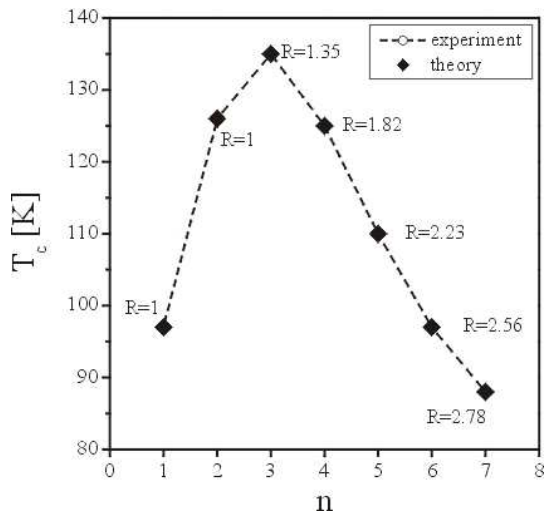}

\caption{Critical temperature $T_{c}$ vs. number of layers $n$ that fits
the experimental data from Ref. \onlinecite{MultiLayer1} and the
calculated values of parameter $R$ describing the ratio of doping
of outer to inner planes. Dashed line is only guide for the eye.\label{Fig2}}
\end{figure}

\begin{figure}
\includegraphics[%
  scale=0.35]{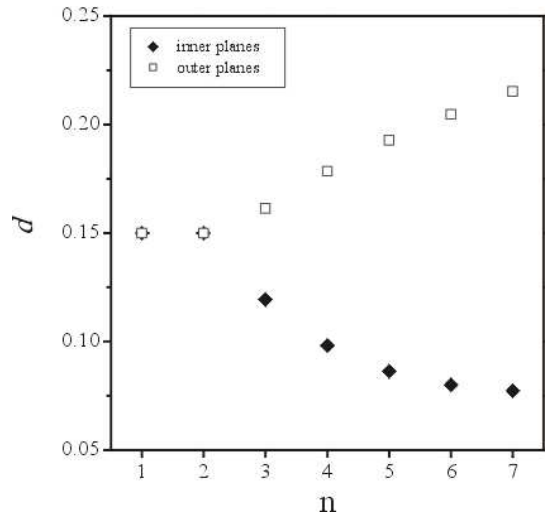}

\caption{Doping of inner $\delta_{ip}$ and outer $\delta_{op}$ planes vs.
number of layers $n$ that corresponds to the data from Fig. \ref{Fig2}.\label{cap:Doping}}
\end{figure}

The obtained values of the ratio $R$ are presented in Fig. \ref{Fig2},
while resulting dopings of inner and outer planes are shown in Fig.
\ref{cap:Doping}. From these results we can infer that for $n=1,\,2,\,3$
the critical temperature is rising due to the increased inter-layer
(intra-block) phase stiffness $J_{\bot}>J_{\bot}^{'}$, while the
charge redistribution for $n=3$ is rather small. For values of $n>3$
the critical temperature is very sensitive to charge redistribution
and we have a decrease from $T_{c}=135K$ for $n=3$ ($R=1.35$) to
$T_{c}=125K$ for $n=4$ ($R=1.82$). The calculated values of $R$
are in good agreement with experimental results based on NMR measurement
of \hbox{HgBa$_2$Ca$_4$Cu$_5$O$_y$} compound by Kotegawa, Ref. \onlinecite{ChargeDistrib}.
Interestingly, the obtained parameters $R$ for $n=3,\,4$ are much
smaller than those found in Ref. \onlinecite{ChargeDistrib2} ($R=4.5$
and $R=9$, respectively) calculated using simplified considerations
based on competition between electrostatic and band energy estimates.
However, in the real materials charge depletion in the inner layers
may be high enough to destroy the superconductivity in these layers,
i.e. doping of one type of planes may be well outside superconducting
region. This scenario have also been confirmed experimentally by NMR
studies of five-layer compound \hbox{HgBa$_2$Ca$_4$Cu$_5$O$_y$}.\cite{AFandSCExp}
Only outer \hbox{Cu-O} planes are superconducting with $T_{c}=108K$,
while inner planes exhibit a static antiferromagnetic ordering with
$T_{N}=60K$. This suggest possible extension of our model to include
independent in-plane phase stiffnesses of the inner and outer layers.

\begin{table}
\begin{tabular}{|c|c|c|c|c|}
\hline 
~~~~$n$~~~~&
$R$ (this work)&
$R$ (Ref. \onlinecite{ChargeDistrib})&
$R$ (Ref. \onlinecite{ChargeDistrib2})&
$R$ (Ref. \onlinecite{AFandSCExp})\tabularnewline
\hline
\hline 
1&
1&
1&
1&
n.a.\tabularnewline
\hline 
2&
1&
1&
1&
n.a.\tabularnewline
\hline 
3&
1.35&
1.14&
4.5$^{a}$&
n.a.\tabularnewline
\hline 
4&
1.82&
1.49&
9$^{b}$&
n.a.\tabularnewline
\hline 
5&
2.23&
1.64&
n.a.&
3.8$^{c}$\tabularnewline
\hline 
6&
2.56&
n.a.&
n.a.&
n.a.\tabularnewline
\hline 
7&
2.78&
n.a.&
n.a.&
n.a.\tabularnewline
\hline
\end{tabular}

\caption{Calculated values of the ratio $R$ (this work, second column) as
a function of number of layers $n$ using the data for $T_{c}\left(n\right)$
from Kuzemskaya, et al. Ref. \onlinecite{MultiLayer1}, compared with
the results from other works: $R$ based on calculated values of charge
concentration of outer (0.45$^{a,b}$) and inner layers (0.1$^{a},$0.05$^{b}$)
using a theory of competition between electrostatic and band energy
estimates (Di Stasio, et al., Ref. \onlinecite{ChargeDistrib2});
nuclear magnetic resonance measurement of \hbox{HgBa$_2$Ca$_4$Cu$_5$O$_y$}
compound$^{c}$ with the values of charge concentration of outer (0.212)
and inner layers (0.057); n.a. -- non-available.}
\end{table}


\section{Summary and conclusions}

We have considered a model of $n$-layer high-temperature cuprates
of homologous series like \hbox{HgBa$_2$Ca$_{n-1}$Cu$_n$O$_{2+2n+\delta }$}
to study the dependence of the critical temperature $T_{c}\left(n\right)$
on the number $n$ of \hbox{Cu-O} planes in the elementary cell.
We focus on the ``phase only'' description of the high-temperature
superconducting system motivated by the experimental evidence that
the ordering of the phase degrees of freedom is responsible for the
emergence of the superconducting state with long-range order. To this
end, we have proposed a three-dimensional semi microscopic XY model
with two-component vectors that involve phase variables and adjustable
parameters representing microscopic phase stiffnesses. The model fully
implements complicated stacked plane structure along $c$-axis to
capture the layered composition of homologous series. This allows
us to go well beyond widely used phenomenological Lawrence-Doniach
model for layered superconductors. Implementing spherical closure
relation we have solved the phase XY model exactly with the help of
transfer matrix method and calculated $T_{c}\left(n\right)$ for chosen
system parameters and arbitrary block size $n$. In this way we were
able to elucidate the role of the $c$-axis anisotropy and its influence
on the critical temperature. Furthermore, by making a physically justified
assumption regarding the doping dependence of the microscopic phase
stiffnesses we were able to calculate the bell shaped curve of $T_{c}\left(n\right)$
as a function of block size $n$ with a maximum at $n=3$ by effectively
accommodating inhomogeneous charge distribution among planes. The
success of our approach suggests to us that at least for the superconducting
phase transition properties studied, one does not need to invoke any
competing ``hidden'' order parameter that appear in other approaches.\cite{Chakravarty}
In the present work, the vanishing of the superconductivity with underdoping
can be understood by the reduction of the in-plane phase stiffnesses
and can be linked to the manifestation of Mott physics (no double
occupancy due to the large Coulomb on-site repulsion) that leads to
the loss of long-range phase coherence while moving toward half-filled
limit ($\delta=0$). For doping $\delta=0$, the fixed electron number
implies large fluctuations in the conjugate phase variable, which
naturally translates into the reduction of the microscopic in-plane
phase stiffnesses and destruction of the superconducting long-range
order in this limit. In the opposite region of large $\delta$, the
onset of a pair-breaking effect (at the pseudogap temperature $T^{*}$)
can deplete the microscopic phase stiffnesses, thus reducing the critical
temperature. Therefore, it would be interesting to relate the phase
stiffnesses parameters used in the present paper with the microscopic
material characteristics of the electronic system (like hopping parameters,
antiferromagnetic exchange and Coulomb energy) to establish a link
between our semi-microscopic approach and physics of strongly correlated
systems. The charge redistribution taking place in multi-layered cuprates
also points out the profound impact of charging effects and unusual
role of the internal electric fields in cuprate superconductors.\cite{Hirsch} 

\begin{acknowledgments}
T. A. Zaleski would like to thank The Foundation for Polish Science
for support.
\end{acknowledgments}
\appendix

\section{Matrix diagonalization\label{sec:Matrix-diagonalization}}

The transfer matrix method requires raising transfer matrix to an
arbitrary power. If matrix $\mathbf{T}$ of the order $m$ has $m$
distinct eigenvalues $\lambda_{1},...,\lambda_{m}$ it can be diagonalized:\begin{equation}
\mathbf{T}=\mathcal{UDU}^{-1},\end{equation}
where $\mathcal{U}$ is a matrix built of eigenvectors $\hat{\varepsilon}_{1},\dots,\hat{\varepsilon}_{m}$
of $\mathbf{T}$:\begin{equation} \mathcal{U}= \left[ \begin{tabular}{|c|c|c|} \cline{1-1} \cline{3-3} & & \\ $\hat{\varepsilon }_1 $ & \dots & $\hat{\varepsilon }_m $ \\ & & \\ \cline{1-1} \cline{3-3} \end{tabular} \right] \end{equation}while
\begin{equation}
\mathcal{D}=\left[\begin{array}{ccc}
\lambda_{1} & 0 & 0\\
0 & \ddots & 0\\
0 & 0 & \lambda_{m}\end{array}\right].\end{equation}
 is diagonal matrix. In result any integer power $p$ of $\mathbf{T}$
can be simply found:\begin{equation}
\mathbf{T}^{p}=\underbrace{\left(\mathcal{UDU}^{-1}\right)...\left(\mathcal{UDU}^{-1}\right)}_{p}=\mathcal{U}\mathcal{D}^{p}\mathcal{U}^{-1},\end{equation}
since $\mathcal{U}^{-1}\mathcal{U}=1$. Finally:\begin{equation}
\mathbf{T}^{p}=\mathcal{U}\left[\begin{array}{ccc}
\lambda_{1}^{p} & 0 & 0\\
0 & \ddots & 0\\
0 & 0 & \lambda_{m}^{p}\end{array}\right]\mathcal{U}^{-1}.\end{equation}
This procedure is utilised in Sec. III. The columns of the matrix
$\mathcal{U}_{1}$ in Eq. \ref{eq:TransMatrixT1} contain eigenvectors
of matrix $\mathbf{T}_{1}$:\begin{widetext}\begin{equation}
\mathcal{U}_{1}=\left[\begin{array}{cc}
{\displaystyle \frac{4\left|B_{n}\right|+\left(\beta J_{\bot}^{'}\right)^{2}\left|B_{n-2}\right|-\sqrt{\Delta}}{8\left|B_{n-1}\right|}} & \,\,\,\,\,{\displaystyle \frac{4\left|B_{n}\right|+\left(\beta J_{\bot}^{'}\right)^{2}\left|B_{n-2}\right|+\sqrt{\Delta}}{8\left|B_{n-1}\right|}}\\
1 & 1\end{array}\right],\end{equation}
where:\begin{equation}
\Delta=\left[16\left|B_{n}\right|+\left(\beta J_{\bot}^{'}\right)^{2}\left|B_{n-2}\right|\right]^{2}-\left(4\beta J_{\bot}^{'}\left|B_{n-1}\right|\right)^{2}\end{equation}
The inverse of the matrix $\mathcal{U}_{1}$ is given by:\begin{equation}
\mathcal{U}_{1}^{-1}=\left[\begin{array}{cc}
{\displaystyle -\frac{4\left|B_{n-1}\right|}{\sqrt{\Delta}}} & {\displaystyle \frac{4\left|B_{n}\right|+\left(\beta J_{\bot}^{'}\right)^{2}\left|B_{n-2}\right|+\sqrt{\Delta}}{2\left|B_{n-1}\right|}}\\
{\displaystyle \frac{4\left|B_{n-1}\right|}{\sqrt{\Delta}}} & {\displaystyle -\frac{4\left|B_{n}\right|+\left(\beta J_{\bot}^{'}\right)^{2}\left|B_{n-2}\right|-\sqrt{\Delta}}{2\left|B_{n-1}\right|}}\end{array}\right].\end{equation}
Similarly, the matrix $\mathcal{U}_{2}$ in Eq. \ref{eq:TransMatrixT2}
is built of eigenvectors of $\mathbf{T}_{2}$:\begin{equation}
\mathcal{U}_{1}=\left[\begin{array}{cc}
\frac{1}{2}\left(\zeta-\frac{\beta J_{\|}}{2}\right)-\frac{1}{2}\sqrt{\left[\left(\zeta-\frac{\beta J_{\|}}{2}\right)^{2}-\left(\beta J_{\bot}\right)^{2}\right]} & \,\,\,\,\,\,\frac{1}{2}\left(\zeta-\frac{\beta J_{\|}}{2}\right)+\frac{1}{2}\sqrt{\left[\left(\zeta-\frac{\beta J_{\|}}{2}\right)^{2}-\left(\beta J_{\bot}\right)^{2}\right]}\\
1 & 1\end{array}\right].\end{equation}
\end{widetext}

\end{document}